\documentclass{article}
\usepackage{iclr2027_conference,times}
\usepackage{subcaption} 
\usepackage{comment}
\usepackage{amsmath,amssymb}
\usepackage{hyperref}
\usepackage{url}
\usepackage{booktabs}
\usepackage{graphicx} 
\usepackage{multirow}
\usepackage{array}
\usepackage{pifont}
\usepackage{wrapfig}
\usepackage{makecell}
\usepackage{tabularx}
\usepackage[table]{xcolor}
\usepackage{algorithm}
\usepackage{algpseudocode}
\usepackage{longtable}
\usepackage{capt-of}

\newcommand{\cmark}{\textcolor{markgreen}{\ding{51}}}
\newcommand{\xmark}{\textcolor{markred}{\ding{55}}}

\newcommand{\ours}{\rowcolor{ourgray}}

\newcolumntype{C}{>{\centering\arraybackslash}p{3.2em}}
\newcolumntype{Y}{>{\centering\arraybackslash}X}

\definecolor{markgreen}{RGB}{34,139,34}
\definecolor{markred}{RGB}{200,40,40}
\definecolor{markorange}{RGB}{230,140,0}

\definecolor{ourgray}{gray}{0.92}

\title{SEW: Style-Encoded Watermarking of LLM-Generated Code}

\author{
Soohan Lim,
Hyundong Jin,
Yo-Sub Han\thanks{Corresponding author.}\\
Yonsei University, Seoul, Republic of Korea\\
   \texttt{\{%
   \href{mailto:aness1219@yonsei.ac.kr}{aness1219},%
   \href{mailto:tuzi04@yonsei.ac.kr}{tuzi04},%
   \href{mailto:emmous@yonsei.ac.kr}{emmous}%
   \}@yonsei.ac.kr}
}

\iclrfinalcopy

\begin{document}
\maketitle

\lhead{Preprint. Under review.}

\begin{abstract}
Code watermarking supports provenance tracking for code generated by LLMs.
Modifying token selection to embed watermarks as an LLM generates
code can create a trade-off between detectability and functional correctness.
Other methods instead watermark completed code using predefined
transformations or trained neural models.
Recurring patterns can make watermark choices predictable across
programs, while treating patterns common in unwatermarked code as
watermark evidence can cause false detections.
We therefore introduce \textbf{SEW},
which embeds and detects watermarks in already generated code through three components:
(i) code style rules collected from style guides and transformation
rules, with style choices determined by a secret key and each program's structural context;
(ii) style-preference calibration, which evaluates watermark evidence
using style probabilities estimated from human-written code; and
(iii) context-aware style aggregation, which combines evidence from
structurally matching locations assigned the same code style choice,
preventing repeated applications of that choice from inflating watermark evidence.
On CodeContests across three LLMs and three programming languages,
SEW achieves a mean relative improvement of 12.44\% in
TPR@FPR5\% over the baselines and is robust to four non-LLM
code-editing attacks, with only a 0.94\% mean relative decrease.
Our code is available at \url{https://github.com/suhanmen/SEW}.
\end{abstract}

\section{Introduction} \label{sec:introduction}
The use of large language models~(LLMs) for code generation creates a need to verify the
provenance of their outputs.
Code watermarking addresses this need by embedding detectable signals
that support subsequent attribution~\citep{lee2024sweet,yang2024srcmarker}.
These signals must distinguish marked code from unmarked code without
compromising program behavior.
A useful code watermark must therefore combine reliable detection with
functional correctness.

Watermarks can be embedded by modifying token selection while an LLM
generates code or by transforming the completed code.
Modifying token probabilities in low-entropy code can create a trade-off
between detectability and functional correctness, motivating entropy-
and syntax-aware strategies~\citep{lee2024sweet,kim2026stone}.
For completed code, existing approaches use fixed transformations or
trained neural embedding and extraction
models~\citep{li2026acw,yang2024srcmarker,zhang2025rosemary}.
Fixed transformations reuse predefined choices, while learned
transformation selectors can also favor recurring patterns.
When the same choices recur across programs, the resulting patterns
can become easier to anticipate and imitate.

We therefore introduce \textbf{SEW}, \textbf{S}tyle-\textbf{E}ncoded
\textbf{W}atermarking, which embeds and detects watermarks in code already generated by an LLM.
SEW integrates three components:
(i) \emph{code style rules}, constructed from style
guides~\citep{pep8,googlestyleguides} and existing code
transformations~\citep{wang2023recode,yang2024srcmarker,li2026acw}.
Each rule supports conversion between two syntax or formatting
variants, with SEW selecting the variant using a secret key and each program's structural context;
(ii) \emph{style-preference calibration}, which evaluates watermark
evidence using probabilities estimated from code style patterns
observed in human-written code, accounting for patterns that also occur without watermarking; and
(iii) \emph{context-aware style aggregation}, which combines evidence
from structurally matching locations assigned the same code style choice, preventing repeated applications of that choice from
inflating watermark evidence.

Table~\ref{tab:requirements} compares SEW and existing code watermarking
approaches regarding model-free detection, pattern resistance, and
functional preservation.
Model-free detection denotes detection without running the generating
LLM or a separately trained extraction model.
Pattern resistance concerns whether an adversary can exploit fixed
or recurring patterns in collected watermarked outputs to infer
watermark choices in other programs.
Functional preservation requires maintaining the pass@1 of the
corresponding unwatermarked outputs under matched evaluation conditions.

\begin{table*}[t]
\centering
\footnotesize
\setlength{\tabcolsep}{8pt}
\renewcommand{\arraystretch}{1.1}
\caption{Comparison of SEW and existing code watermarking.
\cmark: criterion met; \xmark: not met.}
\begin{tabular}{lcccc}
\toprule
Method
  & \makecell{Embedding\\stage}
  & \makecell{Model-free\\detection}
  & \makecell{Pattern-\\resistant}
  & \makecell{Function-\\preserving} \\
\midrule
KGW~\citep{kirchenbauer2023watermark}
  & \multirow{5}{*}{\makecell{During code\\generation}}
  & \cmark & \cmark & \xmark \\
Unigram~\citep{zhao2024unigram}
  & & \cmark & \xmark & \xmark \\
SWEET~\citep{lee2024sweet}
  & & \xmark & \cmark & \xmark \\
STA-1~\citep{mao2025sta1}
  & & \cmark & \cmark & \xmark \\
STONE~\citep{kim2026stone}
  & & \cmark & \cmark & \xmark \\
\midrule
SrcMarker~\citep{yang2024srcmarker}
  & \multirow{4}{*}{\makecell{After code\\generation}}
  & \xmark & \xmark & \xmark \\
RoSeMary~\citep{zhang2025rosemary}
  & & \xmark & \xmark & \xmark \\
ACW~\citep{li2026acw}
  & & \cmark & \xmark & \xmark \\
\ours \textbf{SEW~(ours)}
  & & \cmark & \cmark & \cmark \\
\bottomrule
\end{tabular}
\label{tab:requirements}
\end{table*}

\noindent%
\begin{minipage}[t]{0.49\textwidth}%
\vspace{0pt}%
Figure~\ref{fig:intro} compares TPR@FPR5\% and pass@1
on CodeContests~\citep{abs-2203-07814}, averaged across three LLMs
and three programming languages.
Filled markers denote watermark insertion into completed code,
whereas open markers denote insertion while the LLM generates code.
The dashed line provides the unwatermarked pass@1
reference for assessing functional preservation when watermarking
completed code.
The evaluated baselines have lower detection rates than SEW, and
learning-based transformations of completed code also reduce
pass@1.
SEW achieves a mean TPR@FPR5\% of 99.14\% while preserving
the pass@1 of the corresponding unwatermarked code
across all nine model--language settings.
SEW therefore combines high detectability with preserved benchmark
functional correctness.
\end{minipage}%
\hfill%
\begin{minipage}[t]{0.49\textwidth}%
\vspace{0pt}%
\vspace{-13pt}%
\centering
\includegraphics[width=\linewidth]{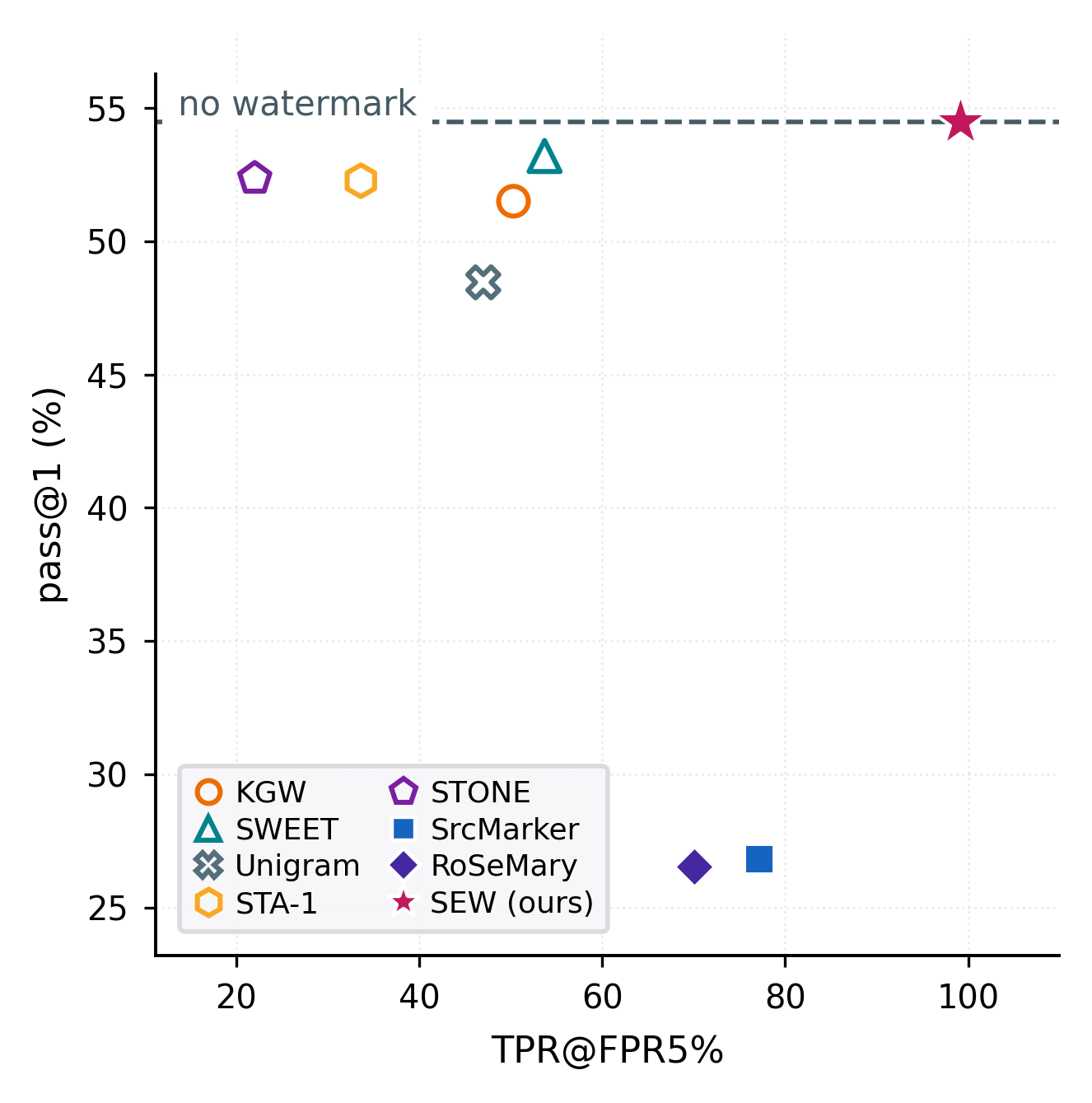}%
\vspace{-10pt}%
\captionof{figure}{Detection vs.\ pass@1.}%
\label{fig:intro}%
\end{minipage}

\vspace{2pt}

We make three main contributions:
\begin{itemize}
    \item \textbf{Structure-aware code style watermarking.}
    We construct a code style rule library and select styles using
    a secret key and structural context without modifying LLM decoding.

    \item \textbf{Calibrated, context-aware style detection.}
    We combine style-preference calibration with context-aware style
    aggregation to evaluate code style patterns as watermark evidence.

    \item \textbf{Empirical validation across models and languages.}
    SEW outperforms baselines in TPR@FPR5\% on CodeContests,
    preserves pass@1, and resists five code-editing attacks.
\end{itemize}

\section{Related Work} \label{sec:related_work}
\paragraph{LLM watermarking.}
Watermarking during LLM generation requires decoding access.
KGW favors keyed green tokens and detects their
overrepresentation~\citep{kirchenbauer2023watermark}, whereas Unigram
uses a fixed vocabulary partition with robustness guarantees under
bounded text edits~\citep{zhao2024unigram}.
STA-1 preserves the unwatermarked next-token distribution in expectation
while limiting the risk of low-probability outputs in low-entropy
contexts~\citep{mao2025sta1}.

\paragraph{Code watermarking.}
Code watermarking accounts for syntax and functional requirements.
SWEET excludes low-entropy positions from embedding and
detection~\citep{lee2024sweet}, whereas STONE restricts embedding
to non-syntactic tokens~\citep{kim2026stone}.
Both require decoding access.
Post-hoc methods transform completed programs.
SrcMarker requires trained neural embedding and extraction
modules~\citep{yang2024srcmarker}.
RoSeMary jointly trains a CodeT5-based embedding module and watermark
extractor~\citep{zhang2025rosemary}.
ACW avoids model training through predefined, idempotent
transformations~\citep{li2026acw}, but its recurring style choices
are susceptible to inference from collected watermarked programs.
SEW instead applies structure-conditioned, secret-key-guided code style
transformations without decoding access or neural model training.

\paragraph{Code style.}
Style documentation~\citep{sunjava1997,pep8, pep515, black, googlestyleguides}
and tool rules describe naming, formatting, and syntactic conventions.
Studies examine coding conventions~\citep{allamanis2014naturalize},
authorship attribution and robustness~\citep{caliskan2015stylometry, quiring2019misleading, li2022ropgen},
style transfer and
generation~\citep{ting2023codestylist, dai2024mpcoder, chen2025abc, zhang2025style2code},
style inconsistencies in LLM-generated code~\citep{wang2025beyond},
and semantics-preserving
transformations~\citep{chakraborty2022natgen, sun2023codemark, li2026acw}.
SEW aims to embed watermarks using context-appropriate syntax and
formatting alternatives without conspicuous stylistic changes.

\begin{figure*}[t]
    \centering
    \includegraphics[width=1\textwidth]{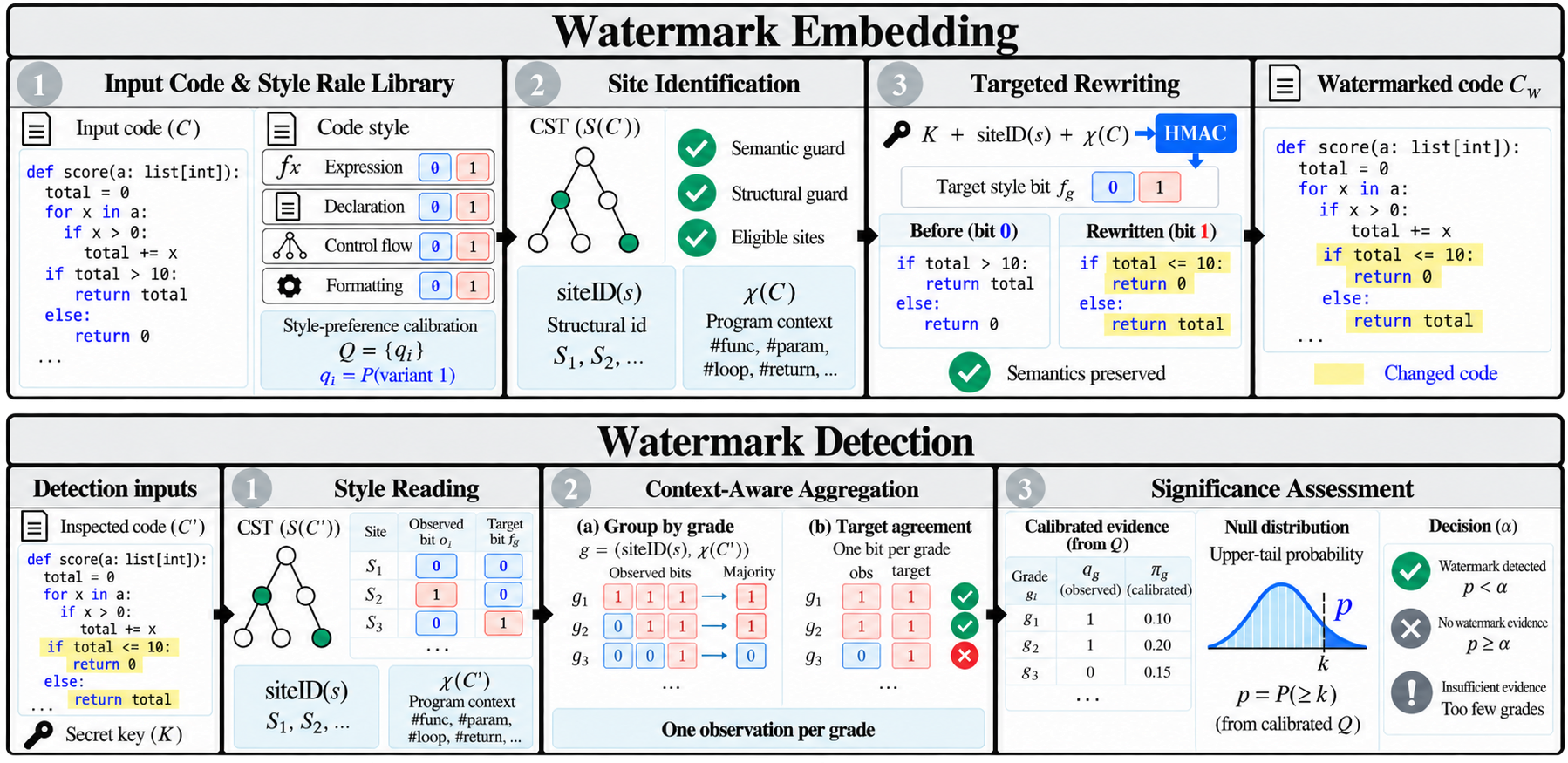}
    \caption{An overview of the SEW pipeline.}
    \label{fig:Overview}
\end{figure*}

\section{Method}\label{sec:method}
Given completed code \(C\) and a secret key \(K\), SEW embeds a
watermark through code style choices and detects it from inspected
code \(C'\) using the same key.
SEW assigns key-dependent style choices according to structural
context and applies them consistently to matching sites within each program.
This design aims to embed watermarks without introducing conspicuous stylistic inconsistencies.
Since these styles can also occur without watermarking, detection
must account for natural style preferences and repeated applications of the same choice.
SEW therefore combines structure-aware style selection with
calibrated, context-aware detection.
(i) The style rule library defines binary syntax and formatting
alternatives, their applicability conditions, and estimated natural variant probabilities.
(ii) Watermark embedding identifies eligible sites and selects
variants using the key and structural context.
(iii) Watermark detection reads the inspected styles, aggregates
repeated choices, and evaluates target agreement against the
estimated natural probabilities.
Neither embedding nor detection requires access to the generation model.
Figure~\ref{fig:Overview} presents the SEW pipeline.

\begin{table}[t]
\centering
\small
\setlength{\tabcolsep}{4pt}
\caption{Code style choices used by SEW, grouped by kind.}
\label{tab:style-rule-examples}
\begin{tabularx}{\linewidth}{@{}l >{\raggedright\arraybackslash}X@{}}
\toprule
\textbf{Kind} & \textbf{Style choices} \\
\midrule
Expression & comparison direction, operand order, redundant parentheses, emptiness and size tests, default arguments, literal forms \\
Declaration & augmented, tuple and chained assignment, multiple declaration, increment form, \texttt{final}/\texttt{const} locals \\
Control flow & branch order, conditional expression, braces on a single statement, loop forms, return forms, comprehensions \\
Formatting & operator/keyword spacing, bracket indent, blank lines, final newline \\
\bottomrule
\end{tabularx}
\end{table}

\subsection{Style Rule Library}\label{ssec:carrier_construction}
The style rule library represents code style alternatives as binary
choices for watermark embedding and detection.
A style pair must support conversion in both directions, but its
applicability and natural occurrence frequencies depend on the code.
We therefore constructed the library through code style collection,
rule-specific applicability conditions, and style-preference calibration.

\paragraph{Code Style Collection.}
We manually collected code style alternatives from seven style guides,
linter and refactoring rule sets, and thirteen prior studies.
We organized these alternatives into Python, Java, and C++ catalogs
covering assignment forms, comparison direction, parenthesization, and operator spacing.
For language \(\ell\), let \(\mathcal{R}_{\ell}\) denote its rule library.
Each rule \(r\in\mathcal{R}_{\ell}\) defines two style variants labeled 0 and 1.
Rule-specific applicability checks cover type, side-effect,
and structural conditions and are shared by embedding and detection.
When applicable, each rule supports bidirectional conversion
to realize the target bit determined by the secret key and
structural context, rather than normalize code toward a single preferred style.
The library contains 29 Python, 22 Java, and 19 C++ rules.
Table~\ref{tab:style-rule-examples} summarizes the rule categories.
Appendix~\ref{app:carrier_rules} provides the full rule list and
examples of applicability conditions.

\paragraph{Style-Preference Calibration.}
The two variants of a style rule need not occur equally often in unwatermarked code.
We estimated their natural frequencies from a separate corpus of user-posted LeetCode solutions~\citep{kaysssleetcode},
without using the CodeContests evaluation solutions.
We applied the fixed rules and their applicability checks to
this corpus and counted eligible occurrences of each variant.
For rule \(r\), let \(n_{r,0}\) and \(n_{r,1}\) denote these counts.
For rules with observed occurrences, the empirical probability of
variant 1 is \(\widehat{q}_r = n_{r,1}/(n_{r,0}+n_{r,1})\).
We obtained \(q_r\) by clipping the estimates to \([0.05,0.95]\),
with \(q_r=1/2\) for rules with fewer than 30 observations or unstable estimates.
The resulting table \(Q=\{q_r\}_{r\in\mathcal{R}_{\ell}}\)
provides the natural-style baseline for detection.
At a site, the estimated chance of matching a target of 1
is \(q_r\), whereas that of matching a target of 0 is \(1-q_r\).
A commonly used target variant can therefore match even without watermarking.
By using its estimated natural probability instead of \(1/2\),
the detector assigns less watermark evidence to such expected matches,
limiting false detections caused by naturally frequent code styles.
Calibration affects detection only; it does not change the
secret-key targets or the embedding procedure.
Appendix~\ref{app:calibration_cost} reports calibration corpus sizes
and probability-estimation times.

\subsection{Watermark Embedding}\label{ssec:watermark_embedding}
Given code \(C\), language \(\ell\), and key \(K\), SEW embeds a
watermark through structure-aware code style choices without modifying
generation.
It analyzes the code to obtain its program context and identifies eligible
style sites from the concrete syntax tree.
For each site, SEW combines \(K\), the program context, and its structural
identifier to determine a binary target style.
Sites with the same structural identifier and context receive the same
target, so SEW applies style choices consistently to structurally
matching locations rather than transforming matching patterns uniformly.

\paragraph{Site Identification.}
SEW parses the code into a concrete syntax tree~(CST) and performs
scope and def-use analysis to identify occurrences of code style rules.
An occurrence enters the eligible site collection \(\mathcal{S}(C)\)
only when it satisfies the rule-specific semantic and structural
conditions for interchanging its two variants.
For example, in Python, the augmented-assignment rule converts between
\texttt{x += y} and \texttt{x = x + y} only when \texttt{x} is
inferred to be numeric; list updates are excluded because in-place
modification differs from reassignment.
The same eligibility checks govern embedding and detection.
Each eligible site \(s\) receives a structural identifier
\(\operatorname{siteID}(s)\) comprising the style rule, statement
type, parent and grandparent node types, and block nesting depth.
This identifier is independent of line numbers and identifier names.
Site selection is public and deterministic; the secret key determines
the target style rather than which sites are selected.

\paragraph{Context-Conditioned Embedding.}
SEW summarizes the program context \(\chi(C)\) using four structural
counts: functions, total parameters, loops, and return statements.
For each eligible site \(s\), SEW combines the secret key \(K\),
the site's structural identifier, and the program context to compute
\[
d_s = \operatorname{HMAC\text{-}SHA256}\bigl(K,\ \operatorname{siteID}(s) \,\Vert\, \chi(C)\bigr),
\qquad
t_s = d_s[0] \bmod 2 \in \{0,1\}.
\]
Here, \(\Vert\) denotes concatenation and \(d_s[0]\) is the first byte
of the HMAC output.
The resulting bit \(t_s\) determines which of the two style variants is used at site \(s\).
SEW reads the current variant and rewrites the site only when its
value differs from \(t_s\).
Sites with the same structural identifier and context receive the
same target under a fixed key, enforcing consistent style choices at
structurally matching locations.
Conversely, the same style rule can receive different targets across
programs with different contexts.
Thus, knowing the public rule alone does not determine its target.

\subsection{Watermark Detection}\label{ssec:statistical_detection}
Given inspected code \(C'\) and the secret key \(K\), SEW reconstructs
the watermark evidence without a reference copy or embedding metadata.
It recovers eligible style sites, recomputes their target bits, and
aggregates repeated observations with the same structural identifier
and program context before statistical testing.
The final test evaluates whether the observed target agreement is
unlikely under calibrated natural style preferences.
It does not use the generation model.

\paragraph{Style Reading and Context-Aware Aggregation.}
SEW applies the same CST parsing, scope and def-use analysis, and
applicability conditions used during embedding to recover \(\mathcal{S}(C')\).
For each recovered site \(s\), it reads the observed style value
\(b_s\in\{0,1\}\) and recomputes \(t_s\) from \(K\),
\(\operatorname{siteID}(s)\), and \(\chi(C')\).
Sites with the same structural identifier and program context are
grouped into a grade \(g=(\operatorname{siteID}(s),\chi(C'))\).
Because all sites in a grade share the same target, counting each
occurrence separately would repeatedly count the same style choice.
Let \(\mathcal{G}\) denote the set of grades and \(t_g\) their shared target.
SEW summarizes the observed variants in each grade by a majority value
\(b_g\), assigning \(b_g=1\) in case of a tie, and records target
agreement as
\[
a_g=\mathbf{1}[b_g=t_g],
\qquad
k=\sum_{g\in\mathcal{G}} a_g.
\]
Each grade therefore contributes one observation, preventing
repeated applications of the same code style choice from inflating
watermark evidence.

\paragraph{Calibrated Statistical Test.}
Code style variants need not occur equally often in unwatermarked code.
SEW therefore uses the calibration table \(Q\) estimated from
human-written code to model target agreement under the null hypothesis.
Using the rule-specific probabilities in \(Q\), let \(\pi_g\) denote
the null probability that grade \(g\) agrees with its target.
Since each grade contributes a binary agreement observation, SEW
models its null agreement as
\(Z_g\sim\operatorname{Bernoulli}(\pi_g)\).
Assuming independent grades, SEW computes the tail probability as
\[
p_{\mathrm{point}}
=
\Pr\left[
\sum_{g\in\mathcal{G}} Z_g \geq k
\right].
\]
This test uses all available grades.
Because formatting rules capture code properties that can vary
independently of program structure, SEW also applies the same test
using only grades from syntax rules, so detection does not rely solely
on formatting-rule evidence.
It combines the resulting \(p_{\mathrm{syn}}\) with
\(p_{\mathrm{point}}\) using a Bonferroni correction, giving
\(p=\min\{1,2\min(p_{\mathrm{point}},p_{\mathrm{syn}})\}\).
Smaller values of \(p\) indicate stronger evidence against the
calibrated null hypothesis.

\section{Experimental Setup}\label{sec:setting}

\subsection{Datasets and Evaluation Metrics}\label{ssec:datasets}
\paragraph{Datasets.}
We used the validation and test splits of
CodeContests~\citep{abs-2203-07814} for Python, Java, and C++.
We refer to unwatermarked model outputs as Vanilla and excluded
problems whose Vanilla output did not complete an evaluable program
within the generation token budget.
Evaluation sets differ across models but are shared across methods
within each model--language setting.
Appendix~\ref{app:denominators} provides the filtering procedure
and problem counts.

\paragraph{Evaluation Metrics.}
We evaluate detection with TPR at 5\% FPR~(TPR@FPR5\%) and AUROC,
which measure performance at a fixed false-positive rate and across
thresholds, respectively.
Functional correctness is measured by pass@1, the fraction of
CodeContests programs that pass all tests.

\subsection{Evaluated Models and Baselines}\label{ssec:baselines}
\paragraph{Evaluated Models.}
We evaluated three open-weight LLMs:
Qwen3.5-9B~\citep{qwen35},
gemma-4-12B-it~\citep{gemmateam2026gemma4}, and
gpt-oss-20b~\citep{openai2025gptoss},
denoted Qwen3.5, gemma-4, and gpt-oss, respectively.
Each generated Python, Java, and C++ solutions, yielding nine
model--language settings.
Qwen3-Coder-30B-A3B-Instruct performed the LLM rewriting attack.

\begin{table*}[t]
\centering
\scriptsize
\setlength{\tabcolsep}{5pt}
\renewcommand{\arraystretch}{0.95}
\caption{
Detection and functional correctness.
T@5: TPR@FPR5\%; P@1: pass@1.
}
\begin{tabular}{l l c *{9}{c}}
\toprule
& & & \multicolumn{3}{c}{\textbf{Python}} & \multicolumn{3}{c}{\textbf{Java}} & \multicolumn{3}{c}{\textbf{C\texttt{++}}} \\
\cmidrule(lr){4-6} \cmidrule(lr){7-9} \cmidrule(lr){10-12}
Model & Method & Type & T@5 & AUROC & P@1 & T@5 & AUROC & P@1 & T@5 & AUROC & P@1 \\
\midrule
& Vanilla     & --       & -- & -- & 41.43 & -- & -- & 36.05 & -- & -- & 36.67 \\
\cmidrule{2-12}
 & KGW         &          & 91.43 & 97.24 & 38.57 & 81.63 & 96.67 & 28.57 & 86.00 & 98.72 & 32.67 \\
 & SWEET       &          & 86.43 & 94.39 & \textbf{50.00} & 79.59 & 91.12 & 27.21 & 86.67 & 94.29 & 32.67 \\
 & Unigram     & Pre-hoc  & 86.43 & 96.05 & 38.57 & 82.31 & 93.65 & 24.49 & 70.00 & 85.51 & 26.67 \\
\multirow{2}{*}{\textbf{Qwen3.5}} & STONE       &          & 45.00 & 67.33 & 45.00 & 23.81 & 66.85 & 25.85 & 20.00 & 43.33 & 31.33 \\
 & STA-1       &          & 75.71 & 93.89 & 48.57 & 51.70 & 85.41 & 26.53 & 83.33 & 96.05 & 34.00 \\
\cmidrule{2-12}
 & SrcMarker   &          & 89.29 & 98.75 & 40.00 & 40.82 & 91.95 & 5.44 & 54.67 & 79.72 & 10.00 \\
 & RoSeMary    & \multirow{2}{*}{Post-hoc} & 98.57 & 98.21 & 40.00 & 40.14 & 94.70 & 5.44 & 80.00 & 89.32 & 8.67 \\
 & ACW         &          & 89.29 & 94.64 & 40.00 & -- & -- & -- & -- & -- & -- \\
\ours \cellcolor{white} & SEW~(ours)  &          & \textbf{100} & \textbf{100} & 41.43 & \textbf{96.60} & \textbf{97.35} & \textbf{36.05} & \textbf{98.67} & \textbf{98.95} & \textbf{36.67} \\
\midrule
& Vanilla     & --       & -- & -- & 87.57 & -- & -- & 78.92 & -- & -- & 81.63 \\
\cmidrule{2-12}
 & KGW         &          & 33.14 & 74.29 & 84.62 & 39.46 & 77.38 & 77.84 & 65.31 & 90.95 & 77.04 \\
 & SWEET       &          & 62.72 & 88.16 & 82.84 & 24.86 & 71.98 & 76.22 & 66.33 & 91.99 & 80.61 \\
 & Unigram     & Pre-hoc  & 68.05 & 95.84 & 84.62 & 24.86 & 71.11 & 72.43 & 64.29 & 92.84 & 77.55 \\
\multirow{2}{*}{\textbf{gemma-4}} & STONE       &          & 33.14 & 64.47 & \textbf{87.57} & 12.97 & 65.77 & \textbf{78.92} & 23.47 & 71.17 & \textbf{82.14} \\
 & STA-1       &          & 7.10 & 33.89 & 82.84 & 3.24 & 51.61 & 77.84 & 9.18 & 36.92 & 80.61 \\
\cmidrule{2-12}
 & SrcMarker   &          & 90.53 & 97.64 & 86.98 & 86.49 & 94.39 & 13.51 & 89.29 & 91.96 & 14.29 \\
 & RoSeMary    & \multirow{2}{*}{Post-hoc} & 97.04 & 97.07 & 85.80 & 25.41 & 95.47 & 13.51 & 90.82 & 95.06 & 14.29 \\
 & ACW         &          & 91.72 & 95.86 & 84.62 & -- & -- & -- & -- & -- & -- \\
\ours \cellcolor{white} & SEW~(ours)  &          & \textbf{100} & \textbf{100} & \textbf{87.57} & \textbf{100} & \textbf{100} & \textbf{78.92} & \textbf{98.98} & \textbf{99.20} & 81.63 \\
\midrule
& Vanilla     & --       & -- & -- & 43.88 & -- & -- & 41.18 & -- & -- & 42.92 \\
\cmidrule{2-12}
 & KGW         &          & 20.41 & 70.14 & 38.78 & 8.82 & 42.87 & 45.10 & 26.61 & 63.94 & 40.34 \\
 & SWEET       &          & 30.61 & 72.98 & \textbf{45.41} & 19.12 & 66.57 & \textbf{46.08} & 27.04 & 75.44 & 37.77 \\
 & Unigram     & Pre-hoc  & 17.86 & 75.02 & 39.29 & 2.45 & 49.34 & 37.75 & 6.87 & 35.03 & 34.76 \\
\multirow{2}{*}{\textbf{gpt-oss}} & STONE       &          & 6.12 & 55.19 & 39.80 & 6.37 & 56.24 & 41.18 & 27.47 & 79.87 & 39.48 \\
 & STA-1       &          & 22.96 & 70.99 & 42.35 & 7.84 & 47.43 & 38.24 & 41.63 & 88.19 & 39.48 \\
\cmidrule{2-12}
 & SrcMarker   &          & 90.82 & 97.18 & 43.37 & 87.25 & 94.40 & 16.67 & 65.67 & 72.98 & 11.16 \\
 & RoSeMary    & \multirow{2}{*}{Post-hoc} & 97.96 & 96.68 & 43.37 & 27.45 & 96.22 & 16.18 & 72.96 & 83.94 & 11.59 \\
 & ACW         &          & 89.29 & 94.64 & 43.37 & -- & -- & -- & -- & -- & -- \\
\ours \cellcolor{white} & SEW~(ours)  &          & \textbf{98.47} & \textbf{98.91} & 43.88 & \textbf{99.51} & \textbf{99.62} & 41.18 & \textbf{100} & \textbf{99.98} & \textbf{42.92} \\
\bottomrule
\end{tabular}
\label{tab:main}
\end{table*}

\paragraph{Baseline Strategies.}
We compare SEW against five token-level watermarking baselines,
KGW~\citep{kirchenbauer2023watermark},
SWEET~\citep{lee2024sweet},
Unigram~\citep{zhao2024unigram},
STONE~\citep{kim2026stone}, and
STA-1~\citep{mao2025sta1}, and three post-hoc baselines,
ACW~\citep{li2026acw},
SrcMarker~\citep{yang2024srcmarker}, and
RoSeMary~\citep{zhang2025rosemary}.
Token-level methods embed watermarks during generation,
whereas post-hoc methods operate on generated code.
ACW was evaluated only in Python because its implementation
supports only that language.
Appendix~\ref{app:implementation} provides details on code generation,
baseline training, and functional evaluation.

\section{Results and Analysis} \label{sec:results}

\subsection{Main Results}\label{ssec:results_main}

Table~\ref{tab:main} reports detection and functional correctness
across nine model--language settings.
SEW achieved the highest TPR@FPR5\% in all nine settings, ranging
from 96.60\% to 100\%, with a mean of 99.14\% and a mean
relative improvement of 12.44\% over the strongest baseline in each setting.
It also achieved the highest AUROC in all nine settings.
Appendix~\ref{app:random_keys} evaluates sensitivity to key choice by reporting the corresponding detection metrics over ten random keys.
SEW also preserved functional correctness, with pass@1 exactly
matching Vanilla in all nine settings.
Post-hoc failures showed that ACW could alter non-commutative string or
list concatenation through operand reordering, while learned SrcMarker
and RoSeMary could change loop behavior by replacing postfix with
prefix decrement in loop conditions.
SEW avoids these failures through rule-specific type, side-effect, and
structural checks that restrict transformations to semantically valid sites.
Token-level methods alter token selection during generation, so pass@1
can rise or fall relative to Vanilla without indicating consistent
functional improvement.
Overall, SEW preserved pass@1 while maintaining high detection performance.

\begin{table*}[t]
\centering
\scriptsize
\setlength{\tabcolsep}{2.5pt}
\renewcommand{\arraystretch}{1.0}
\caption{
TPR@FPR5\% under five code-editing attacks.
AUROC results are in Table~\ref{tab:attack-auroc}.
}
\resizebox{\textwidth}{!}{%
\begin{tabular}{l l *{6}{c} *{6}{c} *{6}{c}}
\toprule
& & \multicolumn{6}{c}{\textbf{Python}} & \multicolumn{6}{c}{\textbf{Java}} & \multicolumn{6}{c}{\textbf{C\texttt{++}}} \\
\cmidrule(lr){3-8} \cmidrule(lr){9-14} \cmidrule(lr){15-20}
Model & Method & None & Ref. & L-d & Cmt. & Ren. & LLM & None & Ref. & L-d & Cmt. & Ren. & LLM & None & Ref. & L-d & Cmt. & Ren. & LLM \\
\midrule
                         & KGW         & 91.43 & 79.29 & 92.14 & 17.86 & 87.86 & 34.29 & 81.63 & 73.47 & 78.91 & 63.27 & 73.47 & 40.82 & 86.00 & 84.00 & 86.00 & 54.00 & 82.67 & 42.67 \\
                         & SWEET       & 86.43 & 82.86 & 87.86 & 12.14 & 85.71 & 31.43 & 79.59 & 80.27 & 79.59 & 24.49 & 75.51 & 25.85 & 86.67 & 84.00 & 86.67 & 40.00 & 80.67 & 33.33 \\
                         & Unigram     & 86.43 & 76.43 & 86.43 & 20.00 & 92.86 & 31.43 & 82.31 & 81.63 & 80.95 & 46.94 & 85.71 & 42.86 & 70.00 & 70.00 & 70.00 & 31.33 & 72.67 & 29.33 \\
                         & STONE       & 45.00 & 42.86 & 45.71 & 6.43 & 49.29 & 26.43 & 23.81 & 23.13 & 24.49 & 10.88 & 23.13 & 8.84 & 20.00 & 20.00 & 20.00 & 9.33 & 22.67 & 9.33 \\
\textbf{Qwen3.5}         & STA-1       & 75.71 & 74.29 & 78.57 & 6.43 & 67.14 & 27.86 & 51.70 & 48.30 & 51.02 & 16.33 & 42.18 & 14.97 & 83.33 & 79.33 & 83.33 & 56.67 & 74.00 & 64.00 \\
\cmidrule{2-20}
                         & SrcMarker   & 89.29 & 90.71 & 88.57 & 89.29 & 7.86 & 10.71 & 40.82 & 40.82 & 39.46 & 40.82 & 10.88 & 2.72 & 54.67 & 54.67 & 54.67 & 54.67 & 38.00 & 0.67 \\
                         & RoSeMary    & 98.57 & 90.00 & 97.86 & 98.57 & 5.00 & 9.29 & 40.14 & 40.14 & 38.10 & 40.14 & 11.56 & 2.72 & 80.00 & 80.00 & 80.00 & 80.00 & 56.67 & 4.67 \\
                         & ACW         & 89.29 & 0.00 & 93.57 & 88.57 & 3.57 & 0.00 & -- & -- & -- & -- & -- & -- & -- & -- & -- & -- & -- & -- \\
\ours \cellcolor{white}  & SEW~(ours)  & \textbf{100} & \textbf{99.29} & \textbf{100} & \textbf{100} & \textbf{100} & \textbf{54.29} & \textbf{96.60} & \textbf{95.92} & \textbf{95.24} & \textbf{96.60} & \textbf{96.60} & \textbf{75.51} & \textbf{98.67} & \textbf{90.67} & \textbf{98.67} & \textbf{98.67} & \textbf{98.67} & \textbf{66.00} \\
\midrule
                         & KGW         & 33.14 & 25.44 & 33.73 & 4.14 & 21.30 & 12.43 & 39.46 & 35.68 & 40.54 & 7.03 & 24.32 & 16.76 & 65.31 & 62.24 & 65.31 & 27.55 & 55.10 & 37.76 \\
                         & SWEET       & 62.72 & 48.52 & 53.85 & 16.57 & 55.62 & 24.85 & 24.86 & 20.00 & 25.41 & 5.41 & 24.86 & 8.65 & 66.33 & 63.78 & 66.33 & 29.08 & 63.27 & 34.69 \\
                         & Unigram     & 68.05 & 60.95 & 68.64 & 8.88 & 50.30 & 18.93 & 24.86 & 23.24 & 24.86 & 8.11 & 20.54 & 8.11 & 64.29 & 52.55 & 64.29 & 13.27 & 51.53 & 21.94 \\
                         & STONE       & 33.14 & 34.32 & 31.36 & 8.28 & 18.93 & 8.88 & 12.97 & 13.51 & 12.97 & 4.32 & 4.32 & 5.41 & 23.47 & 32.65 & 23.47 & 10.71 & 11.22 & 8.67 \\
\textbf{gemma-4}         & STA-1       & 7.10 & 5.92 & 7.10 & 1.18 & 21.30 & 5.33 & 3.24 & 3.24 & 3.24 & 2.70 & 2.16 & 0.54 & 9.18 & 6.63 & 9.18 & 1.02 & 7.65 & 2.04 \\
\cmidrule{2-20}
                         & SrcMarker   & 90.53 & 89.35 & 87.57 & 90.53 & 7.10 & 9.47 & 86.49 & 86.49 & 85.95 & 86.49 & 34.59 & 5.41 & 89.29 & 89.29 & 89.29 & 89.29 & 58.67 & 6.63 \\
                         & RoSeMary    & 97.04 & 86.39 & 95.27 & 97.04 & 5.92 & 5.92 & 25.41 & 25.41 & 24.86 & 25.41 & 9.19 & 1.62 & 90.82 & 90.82 & 90.82 & 90.82 & 56.12 & 4.59 \\
                         & ACW         & 91.72 & 0.00 & 95.27 & 91.72 & 2.37 & 0.00 & -- & -- & -- & -- & -- & -- & -- & -- & -- & -- & -- & -- \\
\ours \cellcolor{white}  & SEW~(ours)  & \textbf{100} & \textbf{99.41} & \textbf{100} & \textbf{100} & \textbf{100} & \textbf{64.50} & \textbf{100} & \textbf{98.92} & \textbf{100} & \textbf{100} & \textbf{100} & \textbf{88.65} & \textbf{98.98} & \textbf{92.86} & \textbf{98.98} & \textbf{98.98} & \textbf{98.98} & \textbf{78.57} \\
\midrule
                         & KGW         & 20.41 & 5.61 & 17.86 & 16.84 & 17.86 & 5.61 & 8.82 & 7.35 & 8.82 & 7.35 & 7.84 & 0.49 & 26.61 & 16.74 & 26.61 & 22.32 & 18.88 & 5.58 \\
                         & SWEET       & 30.61 & 21.94 & 31.63 & 14.29 & 21.43 & 23.47 & 19.12 & 20.10 & 18.63 & 12.75 & 15.69 & 3.43 & 27.04 & 26.61 & 27.04 & 19.31 & 16.74 & 3.86 \\
                         & Unigram     & 17.86 & 14.80 & 20.92 & 13.27 & 12.24 & 17.86 & 2.45 & 2.45 & 2.45 & 2.45 & 2.45 & 0.00 & 6.87 & 9.01 & 6.87 & 6.01 & 6.44 & 1.29 \\
                         & STONE       & 6.12 & 1.02 & 6.12 & 5.10 & 4.59 & 2.55 & 6.37 & 5.88 & 5.88 & 6.37 & 6.86 & 2.45 & 27.47 & 21.03 & 27.47 & 29.18 & 34.33 & 23.61 \\
\textbf{gpt-oss}         & STA-1       & 22.96 & 18.88 & 23.47 & 17.35 & 15.82 & 9.18 & 7.84 & 6.37 & 8.82 & 7.35 & 11.76 & 6.37 & 41.63 & 30.47 & 41.63 & 37.34 & 40.34 & 40.77 \\
\cmidrule{2-20}
                         & SrcMarker   & 90.82 & 90.82 & 90.31 & 90.82 & 9.18 & 8.67 & 87.25 & 87.25 & 86.27 & 87.25 & 30.39 & 7.35 & 65.67 & 65.67 & 65.67 & 65.67 & 41.20 & 9.01 \\
                         & RoSeMary    & 97.96 & 89.80 & 97.45 & 97.96 & 4.59 & 4.08 & 27.45 & 27.45 & 27.45 & 27.45 & 8.33 & 0.49 & 72.96 & 72.96 & 72.96 & 72.96 & 42.92 & 4.29 \\
                         & ACW         & 89.29 & 1.53 & 91.84 & 87.76 & 3.57 & 0.00 & -- & -- & -- & -- & -- & -- & -- & -- & -- & -- & -- & -- \\
\ours \cellcolor{white}  & SEW~(ours)  & \textbf{98.47} & \textbf{93.88} & \textbf{98.47} & \textbf{98.47} & \textbf{98.47} & \textbf{52.55} & \textbf{99.51} & \textbf{99.51} & \textbf{99.51} & \textbf{99.51} & \textbf{99.51} & \textbf{99.02} & \textbf{100} & \textbf{89.70} & \textbf{100} & \textbf{100} & \textbf{100} & \textbf{93.99} \\
\bottomrule
\end{tabular}%
}
\label{tab:attack-tpr}
\end{table*}

\subsection{Robustness to Code-Editing Attacks}\label{ssec:results_attacks}
Table~\ref{tab:attack-tpr} compares supported methods under five
code-editing attacks: formatting~(Ref.), default linting~(L-d),
comment removal~(Cmt.), identifier renaming~(Ren.), and LLM
rewriting~(LLM).
Under the four non-LLM attacks, SEW achieved 98.21\% mean
TPR@FPR5\%, with a 0.94\% mean relative decrease.
KGW and SWEET degraded substantially after comment removal,
consistent with dependence on comment-related token evidence.
SrcMarker and RoSeMary were less affected by comment removal but
sensitive to identifier renaming.
ACW degraded sharply under formatting that can overwrite its fixed
style patterns.
SEW applies code style rules only at eligible sites and draws
watermark evidence from multiple code elements, reducing dependence
on any single element.
Comment removal and identifier renaming left its TPR@FPR5\% unchanged
in all nine settings, while formatting retained 89.70\%--99.51\%.
LLM rewriting with Qwen3-Coder-30B-A3B-Instruct reduced SEW's
TPR@FPR5\% to 52.55\%--99.02\%, averaging 74.79\%,
but SEW remained highest in all nine settings.
SrcMarker and RoSeMary fell to 0.67\%--10.71\% and
0.49\%--9.29\%, respectively, while ACW reached 0\%
in all three Python settings.
SEW embeds watermarks through context-appropriate code style choices,
applying rules only where their semantic and structural conditions are satisfied.
When these choices and their structural context survive rewriting,
they continue to provide watermark evidence.
This is consistent with SEW retaining the highest TPR@FPR5\%
across all nine settings under the evaluated LLM rewriting attack.
Appendix~\ref{app:attack_auroc} reports the corresponding AUROC results,
while Appendix~\ref{app:robust_analysis} analyzes preservation of SEW's
structural evidence under the attacks.

\subsection{Robustness to Targeted Style Flipping}\label{ssec:results_flipping}
\begin{figure}[h]
    \centering
    \includegraphics[width=1\linewidth]{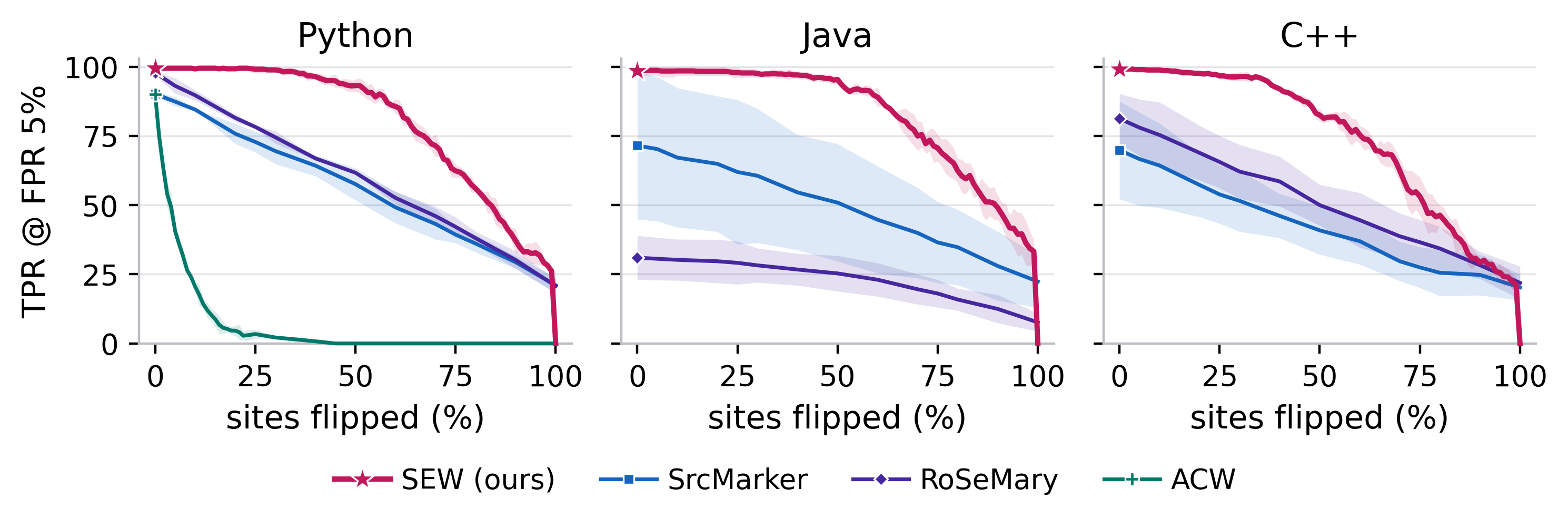}
    \caption{
    TPR@FPR5\% under targeted style flipping; lines show the
    three-model mean and shaded bands the standard deviation.
    }
    \label{fig:flip-tpr5}
\end{figure}

Figure~\ref{fig:flip-tpr5} evaluates a targeted attack in which an
attacker knows each method's watermark rules and reverses a fraction
of its encoded style decisions at their corresponding sites.
Unlike code-editing attacks, this attack targets the
choices used to carry the watermark.
SEW degraded gradually as more sites were reversed, retaining
TPR@FPR5\% of 82.53\%--95.40\% at 50\% flipping and
53.19\%--70.67\% at 75\% flipping across the three languages.
At 75\%, SrcMarker achieved 27.50\%--39.34\% and RoSeMary
18.05\%--42.16\%.
ACW was substantially more sensitive: its TPR@FPR5\% dropped from
90.10\% without attack to 40.57\% at 5\% flipping and
20.49\% at 10\%.
SEW groups repeated sites that share the same structural identifier
and context into a grade and represents them by a single majority
observation.
Thus, when only part of the sites within a grade are flipped, its
watermark evidence can remain unchanged until the majority is reversed.
Complete flipping drives SEW to 0\% because every encoded site is
forced to the opposite target.
SrcMarker and RoSeMary retain nonzero detection even under complete
style flipping. This residual detection comes from identifier-based watermark
components that are not modified by the attack.
Overall, SEW remains detectable even when an attacker directly reverses
a large fraction of its encoded style decisions.

\subsection{Resistance to Rule Inference}\label{ssec:results_rule_inference}
Table~\ref{tab:rule-inference-python} evaluates whether an attacker can
predict watermark decisions after observing \(N\) watermarked programs.
Knowing the rules, the attacker uses observed decisions to predict
style variants on unseen test problems.
Predictions must match the watermarked test programs; unseen decisions
count as incorrect.
We report mean accuracy over ten runs per \(N\);
lower values indicate poorer transfer.
We report Python, the only language supported by all four methods.
Appendix~\ref{app:rule_inference} provides the results for all model--language settings.
The baselines were highly predictable at \(N=10\), reaching
85.97\%--99.67\% accuracy, whereas SEW achieved 19.91\%.
At \(N=100\), SEW reached 45.17\%, while baselines remained above 85\%.
This contrast reflects decision assignment: ACW applies each rule in a
fixed direction, so observed decisions transfer directly to new programs.
SrcMarker and RoSeMary use learned embedding, but their rule-level
transformations recur consistently and remain predictable across programs.
SEW instead derives each target from \(\operatorname{siteID}(s)\) and
program context \(\chi(C)\).
Thus, knowing a rule and its decisions elsewhere does not reveal its
target in a new program; the corresponding
\(\operatorname{siteID}\)--\(\chi(C)\) combination must already have
been observed.
As \(N\) increases, more combinations become known, but unseen ones
still limit prediction when the same rule recurs.

\begin{table}[h]
\centering
\footnotesize
\setlength{\tabcolsep}{4pt}
\renewcommand{\arraystretch}{0.95}
\caption{
Python rule-inference accuracy (\%) over ten runs, averaged across three
models; subscripts denote standard deviations across the models.
}
\begin{tabular}{l *{6}{c}}
\toprule
Method & $N{=}10$ & $N{=}30$ & $N{=}50$ & $N{=}70$ & $N{=}90$ & $N{=}100$ \\
\midrule
SrcMarker        & 90.15$_{\pm 0.72}$ & 90.52$_{\pm 0.87}$ & 90.37$_{\pm 0.69}$ & 90.53$_{\pm 0.70}$ & 90.98$_{\pm 0.65}$ & 90.54$_{\pm 0.43}$ \\
RoSeMary         & 85.97$_{\pm 0.35}$ & 85.65$_{\pm 0.34}$ & 86.16$_{\pm 0.35}$ & 85.69$_{\pm 0.12}$ & 85.96$_{\pm 0.13}$ & 85.87$_{\pm 0.31}$ \\
ACW              & 99.67$_{\pm 0.12}$ & 99.82$_{\pm 0.03}$ & 99.90$_{\pm 0.02}$ & 99.90$_{\pm 0.03}$ & 99.92$_{\pm 0.01}$ & 99.91$_{\pm 0.04}$ \\
\ours SEW~(ours) & \textbf{19.91}$_{\pm 6.42}$ & \textbf{35.83}$_{\pm 7.70}$ & \textbf{39.79}$_{\pm 5.83}$ & \textbf{42.01}$_{\pm 7.14}$ & \textbf{45.63}$_{\pm 3.17}$ & \textbf{45.17}$_{\pm 7.10}$ \\
\bottomrule
\end{tabular}
\label{tab:rule-inference-python}
\end{table}

\subsection{Ablation Study}\label{ssec:results_ablation}
Table~\ref{tab:ablation} isolates the effects of style-preference
calibration and grade-level detection.
Both components improve T@5 without attacks.
For gemma-4 on Java, calibration raises T@5 from 89.19\% to 97.84\%,
while grade-level detection reaches 100\%; for gpt-oss on C++, the
corresponding values are 90.13\%, 99.57\%, and 100\%.
The combined configuration improves over Base in seven of nine
model--language settings, with the other two already at 100\%, but
matches grade-level detection alone in all nine.
Thus, calibration adds no further T@5 improvement after aggregation
at this operating point.
Under attacks, the ranges report language-specific averages over three
models.
Calibration raises Atk. from 73.14\%--73.58\% to
82.38\%--91.15\% and Flip from 31.31\%--34.72\% to
59.38\%--70.95\%.
Grade-level detection yields higher Atk. of 96.34\%--97.69\%; under
Flip, it is stronger in Python and C++, while calibration is stronger
in Java.
Combining both achieves the highest Atk. and Flip in every language,
reaching 97.18\%--99.00\% and 77.49\%--88.00\%, respectively.
Calibration prevents naturally frequent target matches from receiving
excessive evidence, whereas grade-level detection consolidates repeated
sites into a majority observation that is less sensitive to partial
modifications.
Their combination therefore protects detection from both unequal natural
style frequencies and localized modifications, explaining the higher Atk.
and Flip despite unchanged T@5 relative to grade-level detection alone.

\begin{table*}[h]
\centering
\scriptsize
\setlength{\tabcolsep}{4pt}
\renewcommand{\arraystretch}{1.0}
\caption{
Ablating calibration and grade-level detection.
T@5, Atk., and Flip report TPR@FPR5\% without attack,
under four non-LLM editing attacks, and under flipping attacks.
}
\resizebox{\textwidth}{!}{%
\begin{tabular}{l l c c *{9}{c}}
\toprule
& & & & \multicolumn{3}{c}{\textbf{Python}} & \multicolumn{3}{c}{\textbf{Java}} & \multicolumn{3}{c}{\textbf{C\texttt{++}}} \\
\cmidrule(lr){5-7} \cmidrule(lr){8-10} \cmidrule(lr){11-13}
Model & Configuration & Calib & Grade & T@5 & Atk. & Flip & T@5 & Atk. & Flip & T@5 & Atk. & Flip \\
\midrule
 & Base & \xmark & \xmark & \textbf{100} & 74.82 & 35.71 & 95.92 & 77.55 & 38.55 & 94.67 & 74.34 & 35.11 \\
\multirow{2}{*}{\textbf{Qwen3.5}} &  & \cmark & \xmark & \textbf{100} & 86.78 & 70.00 & 95.92 & 90.14 & 73.02 & 97.33 & 81.66 & 64.44 \\
 &  & \xmark & \cmark & \textbf{100} & 99.11 & 77.86 & \textbf{96.60} & 95.41 & 68.25 & \textbf{98.67} & 95.17 & 66.89 \\
\ours \cellcolor{white} & SEW~(ours) & \cmark & \cmark & \textbf{100} & \textbf{99.82} & \textbf{85.71} & \textbf{96.60} & \textbf{96.09} & \textbf{88.89} & \textbf{98.67} & \textbf{96.67} & \textbf{78.00} \\
\midrule
 & Base & \xmark & \xmark & \textbf{100} & 76.18 & 37.08 & 89.19 & 70.41 & 29.19 & 95.41 & 75.13 & 31.63 \\
\multirow{2}{*}{\textbf{gemma-4}} &  & \cmark & \xmark & \textbf{100} & 88.02 & 70.22 & 97.84 & 90.41 & 68.11 & 98.47 & 83.29 & 57.48 \\
 &  & \xmark & \cmark & \textbf{100} & 99.56 & 76.92 & \textbf{100} & 97.43 & 65.77 & \textbf{98.98} & 97.07 & 68.20 \\
\ours \cellcolor{white} & SEW~(ours) & \cmark & \cmark & \textbf{100} & \textbf{99.85} & \textbf{86.39} & \textbf{100} & \textbf{99.73} & \textbf{87.21} & \textbf{98.98} & \textbf{97.45} & \textbf{74.49} \\
\midrule
 & Base & \xmark & \xmark & 96.94 & 69.52 & 27.21 & 91.18 & 72.80 & 36.44 & 90.13 & 69.96 & 27.18 \\
\multirow{2}{*}{\textbf{gpt-oss}} &  & \cmark & \xmark & \textbf{98.47} & 78.83 & 68.20 & \textbf{99.51} & 92.89 & 71.73 & 99.57 & 82.19 & 56.23 \\
 &  & \xmark & \cmark & \textbf{98.47} & 94.39 & 71.43 & \textbf{99.51} & 98.16 & 69.44 & \textbf{100} & 96.78 & 70.96 \\
\ours \cellcolor{white} & SEW~(ours) & \cmark & \cmark & \textbf{98.47} & \textbf{97.32} & \textbf{82.65} & \textbf{99.51} & \textbf{99.51} & \textbf{87.91} & \textbf{100} & \textbf{97.42} & \textbf{79.97} \\
\bottomrule
\end{tabular}%
}
\label{tab:ablation}
\end{table*}

\section{Conclusion}\label{sec:conclusion}
SEW watermarks completed LLM-generated code without model training or
decoding access.
It selects code styles using a secret key and structural context,
combining rule-specific applicability checks, style-preference
calibration, and context-aware style aggregation.
Across three LLMs and three languages on CodeContests, SEW achieved
99.14\% mean TPR@FPR5\% and a 12.44\% mean relative improvement
over each setting's strongest baseline while preserving Vanilla pass@1.
Four non-LLM code-editing attacks yielded a 0.94\% mean relative
TPR@FPR5\% decrease, and SEW resisted cross-program rule inference.
Appendix~\ref{app:limitation} details limited evidence in short code
and reduced coverage for incomplete or structurally invalid code.

\bibliographystyle{iclr2027_conference}
\bibliography{custom}

@article{abs-2203-07814,
  title = {Competition-level code generation with {AlphaCode}},
  author       = {Yujia Li and
                  David H. Choi and
                  Junyoung Chung and
                  Nate Kushman and
                  Julian Schrittwieser and
                  R{\'{e}}mi Leblond and
                  Tom Eccles and
                  James Keeling and
                  Felix Gimeno and
                  Agustin Dal Lago and
                  Thomas Hubert and
                  Peter Choy and
                  Cyprien de Masson d'Autume and
                  Igor Babuschkin and
                  Xinyun Chen and
                  Po{-}Sen Huang and
                  Johannes Welbl and
                  Sven Gowal and
                  Alexey Cherepanov and
                  James Molloy and
                  Daniel J. Mankowitz and
                  Esme Sutherland Robson and
                  Pushmeet Kohli and
                  Nando de Freitas and
                  Koray Kavukcuoglu and
                  Oriol Vinyals},
  journal = {Science},
  volume  = {378},
  number  = {6624},
  pages   = {1092--1097},
  year    = {2022}
}

@misc{kaysssleetcode,
  title = {{LeetCode} Solution Dataset},
  author = {{kaysss}},
  howpublished = {\url{https://huggingface.co/datasets/kaysss/leetcode-problem-solutions}},
  year = {2025}
}

@misc{qwen35,
    title = {{Qwen3.5}: Towards Native Multimodal Agents},
    author       = {{Qwen Team}},
    howpublished = {\url{https://qwen.ai/blog?id=qwen3.5}},
    year = {2026}
}

@misc{gemmateam2026gemma4,
    author       = {Sherif {El Abd} and
                      Vaibhav Aggarwal and
                      Robin Algayres and
                      Alek Andreev and
                      Olivier Bachem and
                      Ian Ballantyne and
                      Cormac Brick and
                      Victor C{\u{a}}rbune and
                      Michelle Casbon and
                      Mayank Chaturvedi and
                      Aditya Chawla and
                      Victor Cotruta and
                      Alice Coucke and
                      Phil Culliton and
                      Robert Dadashi and
                      Lucas Dixon and
                      Mohamed Elhawaty and
                      Utku Evci and
                      Cl{\'{e}}ment Farabet and
                      Johan Ferret and
                      Filippo Galgani and
                      Sertan Girgin and
                      Jean{-}Bastien Grill and
                      Maarten Grootendorst and
                      Jiaxian Guo and
                      Cassidy Hardin and
                      Yanzhang He and
                      Steven M. Hernandez and
                      Omri Homburger and
                      L{\'{e}}onard Hussenot and
                      Ju{-}yeong Ji and
                      Armand Joulin and
                      Aishwarya Kamath and
                      Parnian Kassraie and
                      Olivier Lacombe and
                      Preethi Lahoti and
                      Ga{\"{e}}l Liu and
                      Gus Martins and
                      Luciano Martins and
                      Tatiana Matejovicova and
                      Ramona Merhej and
                      Nikola Momchev and
                      Sneha Mondal and
                      Ryan Mullins and
                      Sindhu Raghuram Panyam and
                      Shreya Pathak and
                      Sarah Perrin and
                      Andr{\'{e}} Susano Pinto and
                      Etienne Pot and
                      Ang{\'{e}}line Pouget and
                      Alexandre Ram{\'{e}} and
                      Sabela Ramos and
                      Douglas Reid and
                      David Rim and
                      Morgane Rivi{\`{e}}re and
                      Karsten Roth and
                      Louis Rouillard and
                      Omar Sanseviero and
                      Pier Giuseppe Sessa and
                      Shane Settle and
                      Danila Sinopalnikov and
                      Sara Smoot and
                      Piotr Stanczyk and
                      Andreas Steiner and
                      Lawrence Stewart and
                      Ilya O. Tolstikhin and
                      Michael Tschannen and
                      Anton Tsitsulin and
                      Nino Vieillard and
                      Renjie Wu and
                      Pingmei Xu and
                      Haichuan Yang and
                      Edouard Yvinec and
                      Biao Zhang and
                      Li Zhang and
                      Joe Zou and
                      Nicolas Aagnes and
                      Abdelrahman Abdelhamed and
                      Jakub Ad{\'{a}}mek and
                      Shivani Agrawal and
                      Shubham Agrawal and
                      Ibrahim Alabdulmohsin and
                      Jean{-}Baptiste Alayrac and
                      Uri Alon and
                      Chandramouli Amarnath and
                      Ankesh Anand and
                      Chrysovalantis Anastasiou and
                      Setareh Ariafar and
                      Fran{\c{c}}ois{-}Xavier Aubet and
                      Kyriakos Axiotis and
                      Federico Barbero and
                      Joelle K. Barral and
                      Alexei Bendebury and
                      Urs Bergmann and
                      Stanley Bileschi and
                      Kat Black and
                      Mathieu Blondel and
                      Sebastian Borgeaud and
                      Arthur Bra{\v{z}}inskas and
                      others},
    title        = {{Gemma} 4 Technical Report},
    howpublished = {arXiv preprint arXiv:2607.02770},
    year         = {2026}
}

@misc{openai2025gptoss,
  title = {{gpt-oss-120b} \& {gpt-oss-20b} Model Card},
  author = {{OpenAI}},
  howpublished = {arXiv preprint arXiv:2508.10925},
  year = {2025}
}

@inproceedings{kirchenbauer2023watermark,
  title = {A Watermark for Large Language Models},
  author       = {John Kirchenbauer and
                  Jonas Geiping and
                  Yuxin Wen and
                  Jonathan Katz and
                  Ian Miers and
                  Tom Goldstein},
  booktitle    = {Proceedings of the 40th International Conference on Machine Learning, {ICML}},
  volume       = {202},
  pages        = {17061--17084},
  year = {2023}
}

@inproceedings{lee2024sweet,
    title = {Who Wrote this Code? Watermarking for Code Generation},
    author       = {Taehyun Lee and
                      Seokhee Hong and
                      Jaewoo Ahn and
                      Ilgee Hong and
                      Hwaran Lee and
                      Sangdoo Yun and
                      Jamin Shin and
                      Gunhee Kim},
    booktitle    = {Proceedings of the 62nd Annual Meeting of the Association for Computational Linguistics, {ACL}},
    pages = {4890--4911},
    year = {2024}
}

@inproceedings{zhao2024unigram,
  title = {Provable Robust Watermarking for {AI}-Generated Text},
  author       = {Xuandong Zhao and
                  Prabhanjan Vijendra Ananth and
                  Lei Li and
                  Yu{-}Xiang Wang},
  booktitle    = {Proceedings of the 12th International Conference on Learning Representations, {ICLR}},
  year = {2024}
}

@inproceedings{kim2026stone,
    title = {Marking Code Without Breaking It: Code Watermarking for Detecting {LLM}-Generated Code},
    author       = {Jungin Kim and
                  Shinwoo Park and
                  Yo{-}Sub Han},
    booktitle    = {Findings of the Association for Computational Linguistics: {EACL} 2026},
    pages = {3990--4002},
    year = {2026}
}

@inproceedings{mao2025sta1,
    title = {Watermarking Large Language Models: An Unbiased and Low-risk Method},
    author       = {Minjia Mao and
                      Dongjun Wei and
                      Zeyu Chen and
                      Xiao Fang and
                      Michael Chau},
    booktitle    = {Proceedings of the 63rd Annual Meeting of the Association for Computational Linguistics, {ACL}},
    pages        = {7939--7960},
    year = {2025}
}

@article{li2026acw,
  title = {Efficient and Universal Watermarking for {LLM}-Generated Code Detection},
  author       = {Boquan Li and
                      Zirui Fu and
                      Mengdi Zhang and
                      Peixin Zhang and
                      Jun Sun and
                      Xingmei Wang},
  journal = {IEEE Transactions on Software Engineering},
  year = {2026}
}

@inproceedings{yang2024srcmarker,
  title = {{SrcMarker}: Dual-Channel Source Code Watermarking via Scalable Code
    Transformations},
  author       = {Borui Yang and
                  Wei Li and
                  Liyao Xiang and
                  Bo Li},
  booktitle    = {Proceedings of the 45th {IEEE} Symposium on Security and Privacy, {SP}},
  pages        = {4088--4106},
  year = {2024}
}

@misc{zhang2025rosemary,
  title = {Robust and Secure Code Watermarking for Large Language Models via
    {ML}/Crypto Codesign},
  author       = {Ruisi Zhang and
                  Neusha Javidnia and
                  Nojan Sheybani and
                  Farinaz Koushanfar},
  howpublished = {arXiv preprint arXiv:2502.02068},
  year = {2025}
}

@inproceedings{wolf2020transformers,
  title = {Transformers: State-of-the-Art Natural Language Processing},
  author       = {Thomas Wolf and
                  Lysandre Debut and
                  Victor Sanh and
                  Julien Chaumond and
                  Clement Delangue and
                  Anthony Moi and
                  Pierric Cistac and
                  Tim Rault and
                  R{\'{e}}mi Louf and
                  Morgan Funtowicz and
                  Joe Davison and
                  Sam Shleifer and
                  Patrick von Platen and
                  Clara Ma and
                  Yacine Jernite and
                  Julien Plu and
                  Canwen Xu and
                  Teven Le Scao and
                  Sylvain Gugger and
                  Mariama Drame and
                  Quentin Lhoest and
                  Alexander M. Rush},
  booktitle    = {Proceedings of the 2020 Conference on Empirical Methods in Natural Language Processing: System Demonstrations, {EMNLP}},
  pages        = {38--45},
  year = {2020}
}

@misc{husain2019codesearchnet,
  title = {{CodeSearchNet} Challenge: Evaluating the State of Semantic Code Search},
  author       = {Hamel Husain and
                  Ho{-}Hsiang Wu and
                  Tiferet Gazit and
                  Miltiadis Allamanis and
                  Marc Brockschmidt},
  howpublished = {arXiv preprint arXiv:1909.09436},
  year = {2019}
}

@inproceedings{allamanis2014naturalize,
  title = {Learning Natural Coding Conventions},
  author       = {Miltiadis Allamanis and
                  Earl T. Barr and
                  Christian Bird and
                  Charles Sutton},
  booktitle    = {Proceedings of the 22nd {ACM} {SIGSOFT} International Symposium on Foundations of Software Engineering, {FSE}},
  pages        = {281--293},
  year = {2014}
}

@inproceedings{caliskan2015stylometry,
  title = {De-anonymizing Programmers via Code Stylometry},
  author       = {Aylin Caliskan-Islam and
                  Richard E. Harang and
                  Andrew Liu and
                  Arvind Narayanan and
                  Clare R. Voss and
                  Fabian Yamaguchi and
                  Rachel Greenstadt},
  booktitle    = {Proceedings of the 24th {USENIX} Security Symposium, {USENIX Security}},
  pages        = {255--270},
  year = {2015}
}

@inproceedings{li2022ropgen,
  title        = {{RoPGen}: Towards Robust Code Authorship Attribution via Automatic Coding
                  Style Transformation},
  author       = {Zhen Li and
                  Qian (Guenevere) Chen and
                  Chen Chen and
                  Yayi Zou and
                  Shouhuai Xu},
  booktitle    = {Proceedings of the 44th {IEEE/ACM} International Conference on Software Engineering, {ICSE}},
  pages        = {1906--1918},
  year = {2022}
}

@inproceedings{quiring2019misleading,
  title        = {Misleading Authorship Attribution of Source Code using Adversarial
                  Learning},
  author       = {Erwin Quiring and
                  Alwin Maier and
                  Konrad Rieck},
  booktitle    = {Proceedings of the 28th {USENIX} Security Symposium, {USENIX Security}},
  pages        = {479--496},
  year = {2019}
}

@inproceedings{sun2023codemark,
  title        = {{CodeMark}: Imperceptible Watermarking for Code Datasets against Neural
                  Code Completion Models},
  author       = {Zhensu Sun and
                  Xiaoning Du and
                  Fu Song and
                  Li Li},
  booktitle    = {Proceedings of the 31st {ACM} Joint European Software Engineering Conference and Symposium on the Foundations of Software Engineering, {ESEC/FSE}},
  pages        = {1561--1572},
  year = {2023}
}

@inproceedings{chakraborty2022natgen,
  title        = {{NatGen}: generative pre-training by ``naturalizing'' source code},
  author       = {Saikat Chakraborty and
                  Toufique Ahmed and
                  Yangruibo Ding and
                  Premkumar T. Devanbu and
                  Baishakhi Ray},
  booktitle    = {Proceedings of the 30th {ACM} Joint European Software Engineering Conference and Symposium on the Foundations of Software Engineering, {ESEC/FSE}},
  pages        = {18--30},
  year = {2022}
}

@inproceedings{ting2023codestylist,
  title        = {{CodeStylist}: {A} System for Performing Code Style Transfer Using Neural
                  Networks},
  author       = {Chih{-}Kai Ting and
                  Karl Munson and
                  Serenity Wade and
                  Anish Savla and
                  Kiran Kate and
                  Kavitha Srinivas},
  booktitle    = {Proceedings of the 37th {AAAI} Conference on Artificial Intelligence, {AAAI}},
  pages        = {16485--16487},
  year = {2023}
}

@article{chen2025abc,
  title        = {{ABC:} Towards a Universal Code Styler through Model Merging},
  author       = {Yitong Chen and
                  Zhiqiang Gao and
                  Chuanqi Shi and
                  Baixuan Li and
                  Miao Gao},
  journal      = {Proceedings of the {ACM} on Programming Languages},
  volume       = {9},
  number       = {{OOPSLA2}},
  pages        = {1512--1540},
  year = {2025}
}

@misc{zhang2025style2code,
  title        = {{Style2Code}: {A} Style-Controllable Code Generation Framework with
                  Dual-Modal Contrastive Representation Learning},
  author       = {Dutao Zhang and
                  Sergey V. Kovalchuk and
                  Yulong He},
  howpublished = {arXiv preprint arXiv:2505.19442},
  year = {2025}
}

@inproceedings{dai2024mpcoder,
  title        = {{MPCoder}: Multi-user Personalized Code Generator with Explicit and
                  Implicit Style Representation Learning},
  author       = {Zhenlong Dai and
                  Chang Yao and
                  WenKang Han and
                  Ying Yuan and
                  Zhipeng Gao and
                  Jingyuan Chen},
  booktitle    = {Proceedings of the 62nd Annual Meeting of the Association for Computational Linguistics, {ACL}},
  pages        = {3765--3780},
  year = {2024}
}

@article{wang2025beyond,
  title        = {Beyond Functional Correctness: Investigating Coding Style Inconsistencies
                  in Large Language Models},
  author       = {Yanlin Wang and
                  Tianyue Jiang and
                  Mingwei Liu and
                  Jiachi Chen and
                  Mingzhi Mao and
                  Xilin Liu and
                  Yuchi Ma and
                  Zibin Zheng},
  journal      = {Proceedings of the {ACM} on Software Engineering},
  volume       = {2},
  number       = {{FSE}},
  pages        = {690--712},
  year = {2025}
}

@misc{pep8,
    title = {{PEP} 8 -- Style Guide for {Python} Code},
    author       = {Guido van Rossum and
                      Barry Warsaw and
                      Alyssa Coghlan},
    howpublished = {\url{https://peps.python.org/pep-0008/}},
    year = {2001}
}

@misc{pep515,
    title = {{PEP} 515 -- Underscores in Numeric Literals},
    author       = {Georg Brandl and
                      Serhiy Storchaka},
    howpublished = {\url{https://peps.python.org/pep-0515/}},
    year = {2016}
}

@misc{googlestyleguides,
    title = {{Google} Style Guides},
    author       = {{Google}},
    howpublished = {\url{https://google.github.io/styleguide/}},
    year = {n.d.}
}

@misc{sunjava1997,
  title = {Code Conventions for the {Java} Programming Language},
  author = {{Sun Microsystems}},
  howpublished = {\url{https://www.oracle.com/java/technologies/javase/codeconventions-contents.html}},
  year = {1999}
}

@misc{black,
    title = {Black: The Uncompromising {Python} Code Formatter},
    author       = {{\L}ukasz Langa and
                      {contributors to Black}},
    howpublished = {\url{https://github.com/psf/black}},
    year = {2026}
}

@inproceedings{wang2023recode,
    title        = {{ReCode}: Robustness Evaluation of Code Generation Models},
    author       = {Shiqi Wang and
                  Zheng Li and
                  Haifeng Qian and
                  Chenghao Yang and
                  Zijian Wang and
                  Mingyue Shang and
                  Varun Kumar and
                  Samson Tan and
                  Baishakhi Ray and
                  Parminder Bhatia and
                  Ramesh Nallapati and
                  Murali Krishna Ramanathan and
                  Dan Roth and
                  Bing Xiang},
    booktitle    = {Proceedings of the 61st Annual Meeting of the Association for Computational Linguistics, {ACL}},
    pages = {13818--13843},
    year = {2023}
}

\clearpage

\appendix

\section{Implementation Details}\label{app:implementation}
We conducted the experiments using three NVIDIA RTX PRO 6000 GPUs.
We generated unwatermarked outputs, denoted as Vanilla, using
Hugging Face Transformers~\citep{wolf2020transformers}
with greedy decoding and a maximum of 8,192 new tokens.
SEW and the post-hoc baselines were applied to these same outputs.
Token-level baselines used the configurations recommended in
their respective papers and implementations.
Functional correctness was evaluated using pass@1 on all test cases
in the public\_tests, private\_tests, and
generated\_tests fields of each CodeContests problem.
A program was considered correct only if it passed all these tests.

For style-preference calibration, we used user-posted LeetCode
solutions.
The calibration corpus contains 11,834 Python, 10,643 Java,
and 11,594 C++ files.
We estimated natural style probabilities from this corpus without
using the CodeContests evaluation solutions.
We trained all language-specific SrcMarker
models for 25 epochs using the authors' implementation.
Our RoSeMary reimplementation was trained
for 20 epochs on the same datasets.
For both methods, the Python and Java models each used 157,851
training functions from the corresponding filtered
CodeSearchNet~\citep{husain2019codesearchnet} dataset.
The models evaluated on C++ were trained on 3,661 C functions
from GitHub-C. Neither method was trained on CodeContests.

\section{Data Construction}\label{app:denominators}
We used the validation and test splits of
CodeContests~\citep{abs-2203-07814}, retaining problems with a canonical
human solution in the target language.
The initial pools contained 203 Python, 223 Java, and 267 C++ problems.
For each model--language setting, we generated unwatermarked
outputs~(Vanilla) with a maximum of 8,192 new tokens.
We excluded problems whose Vanilla outputs exhausted this budget
without completing a code block, including outputs with an unclosed
code fence and those that never began a code block.
This filtering focuses the comparison on watermarking completed code.
Token-level methods can retain detectable signals even in reasoning-only
outputs, whereas post-hoc methods require generated code as input;
detecting signals without generated code falls outside this evaluation scope.
The same Vanilla-based filtering criterion and retained problem set
were used for all methods within each setting, including token-level
methods that generated watermarked code separately.
Table~\ref{tab:dataset_counts} reports the resulting problem counts.
Filtering did not require compilation or test success, and embedding
failures and abstentions within the retained sets remained in the
detection denominator with a score of zero.
Thus, the filtering did not select programs based on whether a
particular watermarking method succeeded.

Inspection of all truncated outputs identified two distinct patterns.
For Qwen3.5-9B and gemma-4-12B-it, reasoning continued within code comments.
Comments accounted for 91.7\%, 89.6\%, and 91.3\% of lines
in truncated Qwen3.5-9B outputs for Python, Java, and C++, respectively.
The corresponding proportions for gemma-4-12B-it were 91.8\%,
89.0\%, and 93.1\%.
Among the 63 truncated Qwen3.5-9B Python outputs, 61 contained at
least 60\% comment lines.
In contrast, gpt-oss-20b exhausted the token budget during reasoning
before opening a code block, leaving little or no generated code.
The excluded outputs therefore included cases where reasoning within
code comments exhausted the allotted token budget before generation
finished, as well as cases where reasoning exhausted the budget before
code generation began.

\begin{table}[h]
\centering
\small
\setlength{\tabcolsep}{8pt}
\renewcommand{\arraystretch}{1.1}
\caption{
CodeContests problem counts before and after model-specific
truncation filtering.
}
\begin{tabular}{lrrr}
\toprule
Problem set & Python & Java & C\texttt{++} \\
\midrule
Before filtering & 203 & 223 & 267 \\
\midrule
Qwen3.5-9B      & 140 & 147 & 150 \\
gemma-4-12B-it  & 169 & 185 & 196 \\
gpt-oss-20b     & 196 & 204 & 233 \\
\bottomrule
\end{tabular}
\label{tab:dataset_counts}
\end{table}

\section{Style-Preference Calibration Cost}\label{app:calibration_cost}
SEW estimates natural variant probabilities by counting eligible
occurrences of each style variant, without neural model training.
The procedure applies the fixed rule library and applicability
checks to the calibration corpus and converts the observed counts
into rule-specific probabilities.
The rules are not learned or modified during calibration.
As shown in Table~\ref{tab:calibration-cost}, estimation took
23.2--24.7 seconds per language on corpora containing
10,643--11,834 LeetCode files.
The probability tables are reused during detection,
so calibration is not repeated for each inspected program.

\begin{table}[h]
\centering
\small
\setlength{\tabcolsep}{8pt}
\caption{
Size of the LeetCode calibration corpus and the time taken to estimate the
rule-specific natural variant probabilities for each language.
}
\begin{tabular}{l r r}
\toprule
Language & Files & Time (s) \\
\midrule
Python          & 11,834 & 24.1 \\
Java            & 10,643 & 23.2 \\
C\texttt{++}    & 11,594 & 24.7 \\
\bottomrule
\end{tabular}
\label{tab:calibration-cost}
\end{table}

\section{AUROC Analysis under Code-Editing Attacks}\label{app:attack_auroc}
Table~\ref{tab:attack-auroc} complements the TPR@FPR5\% results with
AUROC under five code-editing attacks: formatting, default linting,
comment removal, identifier renaming, and LLM rewriting.
All AUROC values are scaled by 100.
Under the four non-LLM attacks, SEW achieved 96.26--100 across
the 36 model--language--attack settings.
Identifier renaming left its AUROC unchanged in all nine settings,
while comment removal changed it by at most 0.01.
This stability is consistent with SEW identifying sites and assigning
targets through structural context rather than identifier names or
comment content.
Formatting had a larger effect, but SEW retained AUROC of
96.79--99.76.

The baselines exhibit attack-specific weaknesses.
For Qwen3.5 on Python, comment removal reduced KGW's AUROC from
97.24 to 71.30 and SWEET's from 94.39 to 59.61,
consistent with dependence on comment-related token evidence.
SrcMarker and RoSeMary retained their AUROC after comment removal,
but identifier renaming reduced RoSeMary's Python AUROC to
48.90--49.62 across the three models.
ACW similarly fell to 50.00--50.77 under formatting,
consistent with a formatter overwriting its fixed style patterns.
These results show that resistance to an edit affecting one code
element does not imply resistance to edits affecting other elements.
SEW retained high AUROC across all four categories in all three
programming languages rather than depending on the preservation of
comments, names, or formatting alone.

LLM rewriting exposed a larger overall loss in detection performance.
SEW retained AUROC of 80.95--99.56, exceeding all supported
post-hoc baselines in every model--language setting and achieving the
highest AUROC among all methods in seven of the nine settings.
SrcMarker reached 43.46--79.94, RoSeMary reached
45.27--53.88, and ACW reached 50.00 in all three Python settings.
The contrast with the preceding attacks shows that robustness to
individual code edits does not fully extend to LLM rewriting.
For SEW, rewriting that changes \(\operatorname{siteID}(s)\) or the
program context \(\chi(C)\) can change the recomputed target even
when a local style variant remains unchanged.
Thus, preserving surface-level choices alone does not ensure that
the detector reconstructs the same watermark evidence.
The results demonstrate partial robustness to the evaluated rewriting
attack, rather than invariance to arbitrary structural changes.

AUROC and TPR@FPR5\% also distinguish overall score separation from
detection at a low false-positive rate.
For gpt-oss on Python, LLM rewriting reduced SEW's AUROC from
98.91 to 89.63, while TPR@FPR5\% fell from 98.47\% to 52.55\%.
A relatively high AUROC therefore does not guarantee high TPR@FPR5\%.
Conversely, under LLM rewriting of Qwen3.5 C++ outputs, STA-1 achieved
higher AUROC than SEW~(90.43 versus 80.95), but lower
TPR@FPR5\%~(64.00\% versus 66.00\%).
Both metrics are therefore needed: AUROC characterizes ranking across
thresholds, while TPR@FPR5\% identifies how much detection remains
at the evaluated operating point.

\begin{table*}[t]
\centering
\scriptsize
\setlength{\tabcolsep}{2.5pt}
\renewcommand{\arraystretch}{1.0}
\caption{
AUROC under five code-editing attacks.
}
\resizebox{\textwidth}{!}{%
\begin{tabular}{l l *{6}{c} *{6}{c} *{6}{c}}
\toprule
& & \multicolumn{6}{c}{\textbf{Python}} & \multicolumn{6}{c}{\textbf{Java}} & \multicolumn{6}{c}{\textbf{C\texttt{++}}} \\
\cmidrule(lr){3-8} \cmidrule(lr){9-14} \cmidrule(lr){15-20}
Model & Method & None & Ref. & L-d & Cmt. & Ren. & LLM & None & Ref. & L-d & Cmt. & Ren. & LLM & None & Ref. & L-d & Cmt. & Ren. & LLM \\
\midrule
                         & KGW         & 97.24 & 86.41 & 97.30 & 71.30 & 95.08 & 70.02 & 96.67 & 93.69 & 96.11 & 93.20 & 93.19 & \textbf{85.92} & 98.72 & \textbf{97.24} & 98.72 & 93.65 & 98.28 & 86.97 \\
                         & SWEET       & 94.39 & 91.42 & 96.38 & 59.61 & 92.99 & 72.20 & 91.12 & 91.37 & 91.30 & 69.43 & 84.96 & 59.57 & 94.29 & 92.94 & 94.29 & 72.12 & 90.08 & 66.26 \\
                         & Unigram     & 96.05 & 86.25 & 95.91 & 65.48 & 97.94 & 74.12 & 93.65 & 93.52 & 92.98 & 81.54 & 96.47 & 82.99 & 85.51 & 85.21 & 85.51 & 58.82 & 89.30 & 53.84 \\
                         & STONE       & 67.33 & 65.17 & 68.04 & 57.10 & 73.16 & 65.75 & 66.85 & 63.73 & 67.88 & 60.15 & 66.52 & 56.27 & 43.33 & 39.73 & 43.33 & 42.92 & 49.83 & 40.93 \\
\textbf{Qwen3.5}         & STA-1       & 93.89 & 89.78 & 94.14 & 67.38 & 85.04 & 73.91 & 85.41 & 83.27 & 86.39 & 66.54 & 75.78 & 67.61 & 96.05 & 94.85 & 96.05 & 88.58 & 91.59 & \textbf{90.43} \\
\cmidrule{2-20}
                         & SrcMarker   & 98.75 & 98.82 & 98.71 & 98.75 & 80.35 & 79.94 & 91.95 & 91.95 & 90.26 & 91.95 & 56.92 & 43.63 & 79.72 & 79.72 & 79.72 & 79.72 & 73.71 & 46.74 \\
                         & RoSeMary    & 98.21 & 97.45 & 98.13 & 98.21 & 48.90 & 49.05 & 94.70 & 94.70 & 92.24 & 94.70 & 66.48 & 45.27 & 89.32 & 89.32 & 89.32 & 89.32 & 84.44 & 52.39 \\
                         & ACW         & 94.64 & 50.00 & 96.79 & 94.29 & 51.79 & 50.00 & -- & -- & -- & -- & -- & -- & -- & -- & -- & -- & -- & -- \\
\ours \cellcolor{white}  & SEW~(ours)  & \textbf{100} & \textbf{99.37} & \textbf{99.86} & \textbf{100} & \textbf{100} & \textbf{83.60} & \textbf{97.35} & \textbf{96.79} & \textbf{96.26} & \textbf{97.35} & \textbf{97.35} & 84.31 & \textbf{98.95} & 97.15 & \textbf{98.95} & \textbf{98.95} & \textbf{98.95} & 80.95 \\
\midrule
                         & KGW         & 74.29 & 64.84 & 74.15 & 49.87 & 59.95 & 53.24 & 77.38 & 75.02 & 78.54 & 58.05 & 62.87 & 55.69 & 90.95 & 89.18 & 90.95 & 77.21 & 86.32 & 79.01 \\
                         & SWEET       & 88.16 & 81.41 & 84.48 & 73.03 & 86.21 & 68.77 & 71.98 & 61.70 & 74.44 & 49.76 & 73.15 & 60.39 & 91.99 & 90.27 & 91.99 & 77.74 & 90.19 & 78.03 \\
                         & Unigram     & 95.84 & 91.94 & 95.79 & 76.49 & 90.29 & 74.08 & 71.11 & 63.85 & 71.01 & 46.30 & 53.74 & 53.58 & 92.84 & 89.46 & 92.84 & 74.89 & 88.65 & 70.25 \\
                         & STONE       & 64.47 & 66.51 & 64.44 & 46.15 & 52.48 & 37.69 & 65.77 & 65.17 & 61.75 & 55.80 & 36.64 & 54.01 & 71.17 & 78.88 & 71.17 & 65.83 & 49.11 & 56.14 \\
\textbf{gemma-4}         & STA-1       & 33.89 & 33.45 & 34.04 & 29.24 & 69.93 & 46.10 & 51.61 & 47.12 & 55.41 & 55.61 & 40.30 & 43.98 & 36.92 & 31.88 & 36.92 & 19.74 & 31.89 & 17.44 \\
\cmidrule{2-20}
                         & SrcMarker   & 97.64 & 97.63 & 97.11 & 97.64 & 76.03 & 75.99 & 94.39 & 94.39 & 94.36 & 94.39 & 63.63 & 47.22 & 91.96 & 91.96 & 91.96 & 91.96 & 84.76 & 54.76 \\
                         & RoSeMary    & 97.07 & 96.50 & 96.98 & 97.07 & 49.32 & 53.88 & 95.47 & 95.47 & 95.26 & 95.47 & 66.05 & 46.97 & 95.06 & 95.06 & 95.06 & 95.06 & 87.30 & 50.71 \\
                         & ACW         & 95.86 & 50.00 & 97.63 & 95.86 & 51.18 & 50.00 & -- & -- & -- & -- & -- & -- & -- & -- & -- & -- & -- & -- \\
\ours \cellcolor{white}  & SEW~(ours)  & \textbf{100} & \textbf{99.40} & \textbf{99.96} & \textbf{100} & \textbf{100} & \textbf{88.90} & \textbf{100} & \textbf{99.76} & \textbf{100} & \textbf{100} & \textbf{100} & \textbf{93.81} & \textbf{99.20} & \textbf{98.10} & \textbf{99.20} & \textbf{99.19} & \textbf{99.20} & \textbf{90.68} \\
\midrule
                         & KGW         & 70.14 & 47.56 & 68.10 & 67.12 & 65.56 & 50.00 & 42.87 & 38.16 & 45.05 & 37.91 & 35.05 & 25.16 & 63.94 & 44.92 & 63.94 & 59.54 & 45.13 & 43.30 \\
                         & SWEET       & 72.98 & 59.63 & 71.56 & 63.31 & 60.49 & 72.60 & 66.57 & 63.27 & 65.45 & 59.08 & 49.42 & 50.26 & 75.44 & 73.16 & 75.44 & 65.52 & 44.44 & 44.77 \\
                         & Unigram     & 75.02 & 70.51 & 76.47 & 69.69 & 64.67 & 73.33 & 49.34 & 50.99 & 48.52 & 48.18 & 34.36 & 40.19 & 35.03 & 49.16 & 35.03 & 30.64 & 22.39 & 19.13 \\
                         & STONE       & 55.19 & 41.03 & 55.33 & 52.06 & 37.88 & 37.25 & 56.24 & 54.63 & 50.11 & 55.73 & 43.38 & 59.63 & 79.87 & 72.40 & 79.87 & 78.55 & 82.37 & 74.10 \\
\textbf{gpt-oss}         & STA-1       & 70.99 & 65.52 & 72.53 & 66.59 & 63.98 & 60.71 & 47.43 & 42.39 & 51.02 & 44.10 & 61.45 & 52.89 & 88.19 & 84.52 & 88.19 & 86.05 & 89.52 & 85.78 \\
\cmidrule{2-20}
                         & SrcMarker   & 97.18 & 97.11 & 97.01 & 97.18 & 73.25 & 74.22 & 94.40 & 94.40 & 93.90 & 94.40 & 62.28 & 43.46 & 72.98 & 72.98 & 72.98 & 72.98 & 65.05 & 45.64 \\
                         & RoSeMary    & 96.68 & 96.00 & 96.44 & 96.68 & 49.62 & 53.28 & 96.22 & 96.22 & 96.15 & 96.22 & 66.56 & 48.21 & 83.94 & 83.94 & 83.94 & 83.94 & 75.26 & 51.42 \\
                         & ACW         & 94.64 & 50.77 & 95.92 & 93.88 & 51.79 & 50.00 & -- & -- & -- & -- & -- & -- & -- & -- & -- & -- & -- & -- \\
\ours \cellcolor{white}  & SEW~(ours)  & \textbf{98.91} & \textbf{97.74} & \textbf{98.86} & \textbf{98.91} & \textbf{98.91} & \textbf{89.63} & \textbf{99.62} & \textbf{99.62} & \textbf{99.62} & \textbf{99.62} & \textbf{99.62} & \textbf{99.56} & \textbf{99.98} & \textbf{98.04} & \textbf{99.98} & \textbf{99.98} & \textbf{99.98} & \textbf{98.45} \\
\bottomrule
\end{tabular}%
}
\label{tab:attack-auroc}
\end{table*}

\section{Resistance to Rule Inference}\label{app:rule_inference}
Tables~\ref{tab:rule-inference-qwen}, \ref{tab:rule-inference-gemma}, and
\ref{tab:rule-inference-gptoss} cover nine model--language settings.
Knowing the rules, the attacker predicted styles in unseen test
programs from \(N\in\{10,30,50,70,90,100\}\) observed watermarked programs.
Predictions had to match watermarked test-code variants; unobserved
decisions counted as errors.
We report ten-run mean accuracy per \(N\), with subscripted standard
deviations; lower is better.
Python, Java, and C++ test counts were 40, 41, and 20 for Qwen3.5;
69, 84, and 82 for gemma-4; and 93, 103, and 74 for gpt-oss.
Methods shared test problems per setting; ACW supports only Python.
SEW achieved the lowest mean accuracy throughout.
At \(N=10\), baselines reached 85.70\%--99.77\%, versus
2.37\%--26.85\% for SEW.
At \(N=100\), baselines remained at 85.53\%--99.95\%, while
SEW reached 39.51\%--53.14\% in Python,
22.78\%--26.48\% in Java, and 17.53\%--35.38\% in C++.
ACW reuses fixed directions; SrcMarker and RoSeMary showed recurring
rule-level preferences despite learned embedding and random messages.
SEW targets depend on the secret key, \(\operatorname{siteID}(s)\),
and program context \(\chi(C)\).
Under a fixed key, targets transfer to matching site--context
combinations, but not necessarily across contexts for the same rule.
Accuracy generally rose with \(N\), consistent with more reusable
combinations; this attack measures decision transfer, not secret-key
recovery or unrestricted target prediction.

\begin{table}[h]
\centering
\small
\setlength{\tabcolsep}{4pt}
\renewcommand{\arraystretch}{1.0}
\caption{
Rule-inference accuracy (\%) on Qwen3.5-9B.
}
\begin{tabular}{l l *{6}{c}}
\toprule
Language & Method & $N{=}10$ & $N{=}30$ & $N{=}50$ & $N{=}70$ & $N{=}90$ & $N{=}100$ \\
\midrule
                                 & SrcMarker   & 90.12$_{\pm 1.19}$ & 90.73$_{\pm 1.53}$ & 90.55$_{\pm 2.04}$ & 90.90$_{\pm 1.40}$ & 91.04$_{\pm 1.70}$ & 90.71$_{\pm 2.16}$ \\
\multirow{2}{*}{\textbf{Python}} & RoSeMary    & 85.85$_{\pm 1.11}$ & 85.34$_{\pm 1.64}$ & 86.54$_{\pm 1.32}$ & 85.79$_{\pm 1.25}$ & 85.84$_{\pm 0.86}$ & 85.53$_{\pm 0.84}$ \\
                                 & ACW         & 99.54$_{\pm 0.30}$ & 99.84$_{\pm 0.15}$ & 99.92$_{\pm 0.08}$ & 99.90$_{\pm 0.08}$ & 99.92$_{\pm 0.09}$ & 99.89$_{\pm 0.06}$ \\
\ours \cellcolor{white}          & SEW~(ours)  & \textbf{14.18}$_{\pm 5.92}$ & \textbf{29.06}$_{\pm 5.56}$ & \textbf{35.90}$_{\pm 5.87}$ & \textbf{36.39}$_{\pm 5.53}$ & \textbf{44.89}$_{\pm 9.77}$ & \textbf{39.51}$_{\pm 9.30}$ \\
\midrule
                                 & SrcMarker   & 94.16$_{\pm 0.86}$ & 94.49$_{\pm 0.35}$ & 94.64$_{\pm 0.49}$ & 94.54$_{\pm 0.42}$ & 94.42$_{\pm 0.43}$ & 94.72$_{\pm 0.41}$ \\
\textbf{Java}                    & RoSeMary    & 94.49$_{\pm 1.16}$ & 95.16$_{\pm 0.46}$ & 95.24$_{\pm 0.66}$ & 95.08$_{\pm 0.72}$ & 95.08$_{\pm 0.32}$ & 94.97$_{\pm 0.42}$ \\
\ours \cellcolor{white}          & SEW~(ours)  & \textbf{6.13}$_{\pm 3.88}$ & \textbf{15.38}$_{\pm 2.88}$ & \textbf{17.20}$_{\pm 2.81}$ & \textbf{17.95}$_{\pm 3.86}$ & \textbf{18.97}$_{\pm 3.69}$ & \textbf{22.78}$_{\pm 4.47}$ \\
\midrule
                                 & SrcMarker   & 93.62$_{\pm 0.82}$ & 93.63$_{\pm 0.61}$ & 93.71$_{\pm 0.41}$ & 93.76$_{\pm 0.50}$ & 94.00$_{\pm 0.71}$ & 93.65$_{\pm 0.71}$ \\
\textbf{C\texttt{++}}            & RoSeMary    & 94.54$_{\pm 0.76}$ & 94.08$_{\pm 0.72}$ & 95.00$_{\pm 0.89}$ & 94.87$_{\pm 0.68}$ & 95.11$_{\pm 0.62}$ & 94.70$_{\pm 0.72}$ \\
\ours \cellcolor{white}          & SEW~(ours)  & \textbf{2.37}$_{\pm 2.02}$ & \textbf{8.03}$_{\pm 5.47}$ & \textbf{11.81}$_{\pm 5.01}$ & \textbf{17.75}$_{\pm 6.08}$ & \textbf{15.62}$_{\pm 7.36}$ & \textbf{19.54}$_{\pm 6.28}$ \\
\bottomrule
\end{tabular}
\label{tab:rule-inference-qwen}
\end{table}

\begin{table}[h]
\centering
\small
\setlength{\tabcolsep}{4pt}
\renewcommand{\arraystretch}{1.0}
\caption{
Rule-inference accuracy (\%) on gemma-4-12B-it.
}
\begin{tabular}{l l *{6}{c}}
\toprule
Language & Method & $N{=}10$ & $N{=}30$ & $N{=}50$ & $N{=}70$ & $N{=}90$ & $N{=}100$ \\
\midrule
                                 & SrcMarker   & 89.45$_{\pm 1.26}$ & 89.57$_{\pm 1.07}$ & 89.61$_{\pm 1.14}$ & 89.72$_{\pm 1.24}$ & 90.30$_{\pm 0.66}$ & 90.06$_{\pm 0.88}$ \\
\multirow{2}{*}{\textbf{Python}} & RoSeMary    & 85.70$_{\pm 0.89}$ & 85.58$_{\pm 0.75}$ & 85.85$_{\pm 1.44}$ & 85.56$_{\pm 0.86}$ & 85.94$_{\pm 0.62}$ & 86.14$_{\pm 0.64}$ \\
                                 & ACW         & 99.77$_{\pm 0.12}$ & 99.79$_{\pm 0.10}$ & 99.88$_{\pm 0.08}$ & 99.93$_{\pm 0.04}$ & 99.91$_{\pm 0.06}$ & 99.88$_{\pm 0.07}$ \\
\ours \cellcolor{white}          & SEW~(ours)  & \textbf{26.85}$_{\pm 7.16}$ & \textbf{44.20}$_{\pm 4.24}$ & \textbf{46.49}$_{\pm 3.10}$ & \textbf{50.04}$_{\pm 4.22}$ & \textbf{49.10}$_{\pm 2.27}$ & \textbf{53.14}$_{\pm 4.31}$ \\
\midrule
                                 & SrcMarker   & 94.33$_{\pm 0.78}$ & 94.39$_{\pm 0.91}$ & 94.71$_{\pm 0.33}$ & 94.66$_{\pm 0.39}$ & 94.83$_{\pm 0.34}$ & 94.62$_{\pm 0.28}$ \\
\textbf{Java}                    & RoSeMary    & 94.31$_{\pm 1.05}$ & 94.68$_{\pm 1.03}$ & 95.08$_{\pm 0.20}$ & 95.18$_{\pm 0.34}$ & 94.93$_{\pm 0.36}$ & 95.10$_{\pm 0.44}$ \\
\ours \cellcolor{white}          & SEW~(ours)  & \textbf{5.72}$_{\pm 2.28}$ & \textbf{11.37}$_{\pm 2.44}$ & \textbf{15.81}$_{\pm 2.69}$ & \textbf{20.40}$_{\pm 2.75}$ & \textbf{24.50}$_{\pm 2.90}$ & \textbf{24.08}$_{\pm 3.71}$ \\
\midrule
                                 & SrcMarker   & 93.26$_{\pm 0.59}$ & 93.55$_{\pm 0.38}$ & 93.17$_{\pm 0.61}$ & 93.49$_{\pm 0.35}$ & 93.62$_{\pm 0.23}$ & 93.57$_{\pm 0.22}$ \\
\textbf{C\texttt{++}}            & RoSeMary    & 94.29$_{\pm 1.09}$ & 94.57$_{\pm 0.79}$ & 95.00$_{\pm 0.29}$ & 94.84$_{\pm 0.33}$ & 94.85$_{\pm 0.34}$ & 94.82$_{\pm 0.31}$ \\
\ours \cellcolor{white}          & SEW~(ours)  & \textbf{4.66}$_{\pm 1.86}$ & \textbf{9.23}$_{\pm 1.48}$ & \textbf{12.54}$_{\pm 2.27}$ & \textbf{15.76}$_{\pm 2.54}$ & \textbf{18.82}$_{\pm 2.57}$ & \textbf{17.53}$_{\pm 1.62}$ \\
\bottomrule
\end{tabular}
\label{tab:rule-inference-gemma}
\end{table}

\begin{table}[h]
\centering
\small
\setlength{\tabcolsep}{4pt}
\renewcommand{\arraystretch}{1.0}
\caption{
Rule-inference accuracy (\%) on gpt-oss-20b.
}
\begin{tabular}{l l *{6}{c}}
\toprule
Language & Method & $N{=}10$ & $N{=}30$ & $N{=}50$ & $N{=}70$ & $N{=}90$ & $N{=}100$ \\
\midrule
                                 & SrcMarker   & 90.88$_{\pm 1.57}$ & 91.27$_{\pm 1.37}$ & 90.96$_{\pm 1.65}$ & 90.96$_{\pm 1.27}$ & 91.60$_{\pm 0.86}$ & 90.86$_{\pm 1.58}$ \\
\multirow{2}{*}{\textbf{Python}} & RoSeMary    & 86.36$_{\pm 0.87}$ & 86.02$_{\pm 0.66}$ & 86.08$_{\pm 0.64}$ & 85.71$_{\pm 0.83}$ & 86.09$_{\pm 0.80}$ & 85.95$_{\pm 0.64}$ \\
                                 & ACW         & 99.70$_{\pm 0.21}$ & 99.83$_{\pm 0.13}$ & 99.90$_{\pm 0.07}$ & 99.88$_{\pm 0.06}$ & 99.93$_{\pm 0.04}$ & 99.95$_{\pm 0.03}$ \\
\ours \cellcolor{white}          & SEW~(ours)  & \textbf{18.69}$_{\pm 6.08}$ & \textbf{34.23}$_{\pm 4.01}$ & \textbf{36.98}$_{\pm 4.19}$ & \textbf{39.61}$_{\pm 2.63}$ & \textbf{42.89}$_{\pm 3.83}$ & \textbf{42.85}$_{\pm 4.51}$ \\
\midrule
                                 & SrcMarker   & 94.53$_{\pm 0.77}$ & 94.82$_{\pm 0.36}$ & 94.84$_{\pm 0.31}$ & 94.71$_{\pm 0.37}$ & 94.56$_{\pm 0.19}$ & 94.82$_{\pm 0.20}$ \\
\textbf{Java}                    & RoSeMary    & 94.56$_{\pm 1.33}$ & 95.17$_{\pm 0.20}$ & 95.17$_{\pm 0.20}$ & 95.15$_{\pm 0.22}$ & 95.19$_{\pm 0.31}$ & 95.09$_{\pm 0.23}$ \\
\ours \cellcolor{white}          & SEW~(ours)  & \textbf{5.00}$_{\pm 1.38}$ & \textbf{13.25}$_{\pm 2.82}$ & \textbf{18.99}$_{\pm 3.34}$ & \textbf{22.87}$_{\pm 4.06}$ & \textbf{26.74}$_{\pm 1.31}$ & \textbf{26.48}$_{\pm 2.70}$ \\
\midrule
                                 & SrcMarker   & 93.23$_{\pm 0.71}$ & 93.47$_{\pm 0.83}$ & 93.54$_{\pm 0.25}$ & 93.44$_{\pm 0.42}$ & 93.32$_{\pm 0.66}$ & 93.63$_{\pm 0.32}$ \\
\textbf{C\texttt{++}}            & RoSeMary    & 94.79$_{\pm 0.82}$ & 94.79$_{\pm 0.29}$ & 95.15$_{\pm 0.32}$ & 94.99$_{\pm 0.34}$ & 94.91$_{\pm 0.47}$ & 95.12$_{\pm 0.37}$ \\
\ours \cellcolor{white}          & SEW~(ours)  & \textbf{12.58}$_{\pm 3.56}$ & \textbf{22.46}$_{\pm 2.34}$ & \textbf{26.30}$_{\pm 4.50}$ & \textbf{32.41}$_{\pm 4.70}$ & \textbf{34.83}$_{\pm 3.56}$ & \textbf{35.38}$_{\pm 4.29}$ \\
\bottomrule
\end{tabular}
\label{tab:rule-inference-gptoss}
\end{table}


\section{Robustness to the Choice of Secret Key}\label{app:random_keys}
Table~\ref{tab:random-keys} evaluates whether SEW's detection
performance depends on the choice of secret key.
Across ten random 128-bit keys and nine model--language settings,
mean TPR@FPR5\% ranged from 96.60\% to 100\%, while mean AUROC
ranged from 97.43 to 100.
Averaged equally across the nine settings, TPR@FPR5\% was 99.24\%
and AUROC was 99.45.
Seven of the nine settings achieved a mean TPR@FPR5\% of at least
99\%, and seven achieved a mean AUROC of at least 99.
These setting-level results summarize performance over 90
key--setting evaluations.

Variation across keys was limited.
Five of the nine settings had a reported TPR@FPR5\% standard
deviation of 0.00, and the remaining four had standard deviations
between 0.17 and 0.81.
The largest variation occurred for gpt-oss on Python, with standard
deviations of 0.81 for TPR@FPR5\% and 0.51 for AUROC.
Even in this setting, the corresponding means remained 99.23\%
and 99.53, respectively.
Qwen3.5 on Java had the lowest mean TPR@FPR5\% of 96.60\% and
the lowest mean AUROC of 97.43, but its reported standard deviations
were 0.00 and 0.07.
The lower performance in this setting was therefore consistent across
the sampled keys and evident in both reported metrics rather than
driven by an outlying key.

Using a 128-bit key alone does not guarantee stable detection across
key choices.
Changing the key changes the HMAC output and can reassign the target
style for each combination of structural identifier and program
context.
The calibration table remains fixed across keys, but the key-dependent
target determines whether each grade uses \(q_r\) or \(1-q_r\) as its
null agreement probability.
Despite these changes, SEW retained similar detection performance
across the ten sampled keys in all evaluated models and languages.
Style-preference calibration uses variant probabilities estimated from
the calibration corpus, while context-aware style aggregation prevents
repeated occurrences of one target decision from dominating the evidence.
These results provide empirical evidence that SEW's detection
performance is stable across the ten sampled keys rather than tied
to the particular key used in the main evaluation.

\begin{table}[h]
\centering
\small
\setlength{\tabcolsep}{5pt}
\renewcommand{\arraystretch}{1.1}
\caption{
SEW detection over ten random 128-bit keys.
T@5: TPR@FPR5\%.
Entries report the mean over the ten keys with subscripted standard
deviations.
}
\begin{tabular}{l *{6}{c}}
\toprule
& \multicolumn{2}{c}{\textbf{Python}} & \multicolumn{2}{c}{\textbf{Java}} & \multicolumn{2}{c}{\textbf{C\texttt{++}}} \\
\cmidrule(lr){2-3} \cmidrule(lr){4-5} \cmidrule(lr){6-7}
Model & T@5 & AUROC & T@5 & AUROC & T@5 & AUROC \\
\midrule
\textbf{Qwen3.5} & 100$_{\pm 0.00}$   & 100$_{\pm 0.01}$ & 96.60$_{\pm 0.00}$ & 97.43$_{\pm 0.07}$ & 98.47$_{\pm 0.32}$ & 98.97$_{\pm 0.07}$ \\
\textbf{gemma-4} & 100$_{\pm 0.00}$   & 100$_{\pm 0.00}$    & 99.95$_{\pm 0.17}$ & 99.97$_{\pm 0.04}$ & 99.44$_{\pm 0.29}$ & 99.57$_{\pm 0.22}$ \\
\textbf{gpt-oss} & 99.23$_{\pm 0.81}$ & 99.53$_{\pm 0.51}$  & 99.51$_{\pm 0.00}$ & 99.62$_{\pm 0.01}$ & 100$_{\pm 0.00}$   & 99.98$_{\pm 0.02}$ \\
\bottomrule
\end{tabular}
\label{tab:random-keys}
\end{table}

\section{Structural Analysis of Robustness to Code-Editing Attacks} \label{app:robust_analysis}
Attack robustness depends on whether an edit preserves the code elements
from which a detector reconstructs watermark evidence.
Context-dependent token methods such as KGW and SWEET derive evidence
from the exact token sequence and preceding-token-conditioned green
lists, so comment removal or token regeneration can disrupt their signals.
For Qwen3.5 on Python, comment removal reduced TPR@FPR5\% from
91.43\% to 17.86\% for KGW and from 86.43\% to 12.14\% for SWEET.
SrcMarker and RoSeMary include identifier-based watermark components;
identifier renaming reduced their TPR@FPR5\% to 7.10\%--58.67\% and 4.59\%--56.67\%, respectively.
ACW uses fixed style assignments that formatting can overwrite,
reducing its TPR@FPR5\% to 0.00\%--1.53\%.
SEW instead indexes each grade by the structural identifier
\(\operatorname{siteID}(s)\) and program context \(\chi(C)\).
Neither representation includes identifier names, line numbers, or comments.
Surface edits can therefore alter token sequences, names, or formatting
without necessarily changing the structural tuples from which SEW
reconstructs its evidence.

Table~\ref{tab:attack-survival} follows the attack abbreviations in
Table~\ref{tab:attack-tpr} and measures preservation of SEW's structural
evidence locations.
We applied each attack to the watermarked programs and recorded whether
\(\chi(C)\) remained unchanged and what fraction of the pre-attack
grades \((\operatorname{siteID}(s),\chi(C))\) was recovered with the
same tuple afterward.
Across the four non-LLM attacks, \(\chi(C)\) was preserved in
98.47\%--100\% of programs, while 97.19\%--100\% of grades survived
across all model--language settings.
Although comment removal changed 26\%--37\% of tokens and identifier
renaming changed 10\%--13\%, both preserved nearly all structural
evidence locations.
The detector reconstructs targets from these surviving structural tuples
and separately tests syntax-rule grades, retaining watermark evidence
when comments, identifiers, or formatting are modified.
This design supports SEW's high detection performance across all four
non-LLM attacks.

LLM rewriting is the broadest evaluated attack because it regenerates
tokens, can rename identifiers, and normalizes formatting in one step.
These edits jointly disrupt the token sequences, identifiers, and fixed
style patterns used by the baselines.
Under Qwen3-Coder-30B-A3B-Instruct, SrcMarker, RoSeMary, and ACW fell
to 0.00\%--10.71\%, while the strongest token-level baseline in each
setting reached at most 64.00\%.
SEW retained TPR@FPR5\% of 52.55\%--99.02\% and remained highest in all nine settings.
SEW reconstructs each target from the secret key,
\(\operatorname{siteID}(s)\), and \(\chi(C)\), so token regeneration
or identifier changes do not necessarily erase the corresponding
evidence location.

The survival rates also clarify the model- and language-specific outcomes.
Program context remained unchanged in 77.14\%--100\% of programs, and
70.77\%--99.76\% of grades survived rewriting.
In Java and C++, 83.7\%--86.7\% of surviving grades that agreed with
their targets before rewriting remained target-agreeing afterward.
In Python, formatting normalization toward PEP~8 reduced the
corresponding rate to 63.9\%.
For Qwen3.5, \(\chi(C)\) changed in 20.95\%--22.86\% of programs,
causing every target in each affected program to be recomputed from a
different HMAC input.
Despite target reassignment in these programs and language-specific
style normalization, SEW retained the highest TPR@FPR5\% in all nine settings.
These results show that SEW reconstructs usable watermark evidence from
the structural tuples that survive rewriting, allowing it to outperform
every evaluated baseline across all nine settings.

\begin{table*}[h]
\centering
\scriptsize
\setlength{\tabcolsep}{3.2pt}
\renewcommand{\arraystretch}{1.05}
\caption{
SEW context preservation~(Ctx) and grade survival~(Grd) under
code-editing attacks (\%).}
\label{tab:attack-survival}
\begin{tabular*}{\textwidth}{@{\hspace{8pt}\extracolsep{\fill}}cc *{10}{c}@{\hspace{8pt}}}
\toprule
& &
\multicolumn{2}{c}{Ref.} &
\multicolumn{2}{c}{L-d} &
\multicolumn{2}{c}{Cmt.} &
\multicolumn{2}{c}{Ren.} &
\multicolumn{2}{c}{LLM} \\
\cmidrule(lr){3-4}
\cmidrule(lr){5-6}
\cmidrule(lr){7-8}
\cmidrule(lr){9-10}
\cmidrule(lr){11-12}
Language & Model &
Ctx & Grd & Ctx & Grd & Ctx & Grd &
Ctx & Grd & Ctx & Grd \\
\midrule
\multirow{3}{*}{\textbf{Python}}
& Qwen3.5
& 99.29 & 98.76 & 99.29 & 97.19 & 100 & 99.77
& 100 & 99.95 & 77.14 & 70.77 \\
& gemma-4
& 99.41 & 99.32 & 100 & 98.72 & 100 & 99.90
& 100 & 99.98 & 88.76 & 84.61 \\
& gpt-oss
& 100 & 99.35 & 100 & 98.20 & 98.47 & 99.27
& 100 & 100 & 98.98 & 98.93 \\
\midrule
\multirow{3}{*}{\textbf{Java}}
& Qwen3.5
& 100 & 99.35 & 98.59 & 97.26 & 100 & 99.51
& 100 & 100 & 78.87 & 74.58 \\
& gemma-4
& 100 & 99.15 & 100 & 99.40 & 100 & 99.80
& 100 & 100 & 92.43 & 89.30 \\
& gpt-oss
& 100 & 99.14 & 100 & 99.67 & 100 & 99.86
& 100 & 100 & 100 & 99.76 \\
\midrule
\multirow{3}{*}{\textbf{C\texttt{++}}}
& Qwen3.5
& 100 & 99.12 & 100 & 100 & 100 & 99.53
& 100 & 100 & 79.05 & 73.47 \\
& gemma-4
& 100 & 98.95 & 100 & 100 & 100 & 99.65
& 100 & 100 & 90.82 & 84.87 \\
& gpt-oss
& 100 & 98.73 & 100 & 100 & 100 & 99.72
& 100 & 100 & 99.57 & 99.02 \\
\bottomrule
\end{tabular*}
\end{table*}

\section{Limitations}\label{app:limitation}
SEW supports provenance tracking through code style watermarking,
but its applicability depends on the available style evidence
and the structural validity of generated code.

\paragraph{Limited evidence in short programs.}
Very short programs can contain too few eligible style sites
for reliable watermark detection.
Rule-specific applicability conditions further restrict embedding sites.
Sites sharing the same structural identifier and program context
contribute only one grade-level observation, so repetitions do not
add observations to the statistical test.
Even functionally correct programs can therefore provide
insufficient watermark evidence.

\paragraph{Coverage of incomplete or invalid code.}
As a post-hoc method, SEW requires completed code that can be
parsed and rewritten.
Incomplete or structurally invalid outputs can prevent site
identification and style transformations, limiting watermark coverage.
These input defects precede embedding and are not caused by
watermark insertion.
Detection performance on supported inputs must therefore be
distinguished from watermark coverage across all generated outputs.

\section{Code Style Rules and Applicability Conditions}
\label{app:carrier_rules}
Table~\ref{tab:rules-full} lists the code style rules used by SEW:
29 for Python, 22 for Java, and 19 for C++.
Each rule defines two variants labeled 0 and 1, with
bidirectional conversion subject to rule-specific applicability conditions.
SEW selects the target variant using the secret key and structural
context rather than enforcing a fixed style across programs.
The library comprises four categories.
Expression rules cover comparison direction, operand order,
parenthesization, tests, and literal forms.
Declaration rules cover assignment and declaration forms, increments,
and local modifiers.
Control-flow rules cover branch order, conditional expressions,
loop and return forms, and comprehensions.
Formatting rules cover spacing, indentation, blank lines, and the final newline.
The first three categories contain syntax rules, whereas formatting
rules modify only whitespace and layout.

A matching code pattern alone does not make an occurrence eligible:
SEW also checks rule-specific type, side-effect, and structural conditions.
For example, Python augmented-assignment conversion requires the
assignment target to be inferred as numeric.
List updates are excluded because in-place modification differs 
from reassignment.
Addition-operand exchange also requires numeric operands, excluding
order-sensitive string, list, and tuple concatenation.
Comparison-direction changes require side-effect-free operands
because exchanging their positions can change evaluation order.
Branch-order changes apply to paired \texttt{if}--\texttt{else}
branches and exclude \texttt{elif} chains.
Embedding and detection share these checks, so both identify eligible
sites using the same applicability criteria.
The table presents representative variants in one supported language;
shared rules use corresponding language-specific forms.
Length and emptiness tests use operations such as \texttt{len(x)},
\texttt{v.size()}, and \texttt{c.isEmpty()}.
The default-argument rule pairs \texttt{range(n)} with
\texttt{range(0, n)} in Python and \texttt{s.indexOf(c)} with
\texttt{s.indexOf(c, 0)} in Java.
Local modifiers use \texttt{final} in Java and \texttt{const} in C++.
In Python, the infinite-loop variants are \texttt{while True:} and
\texttt{while 1:}, while conditional expressions use
\texttt{x = a if c else b}.
Italicized entries describe transformation patterns rather than
executable code.

\begin{table}[h]
\centering
\footnotesize
\setlength{\tabcolsep}{3pt}
\renewcommand{\arraystretch}{1.05}
\caption{
SEW code style rules and variants.
Py: Python; All: Python, Java, and C\texttt{++}.
}
\label{tab:rules-full}
\begin{tabularx}{\linewidth}{@{}c >{\raggedright\arraybackslash}X l l l@{}}
\toprule
& \textbf{Style choice} & \textbf{Variant 0} & \textbf{Variant 1} & \textbf{Lang.} \\
\midrule
\multirow{33}{*}{\rotatebox[origin=c]{90}{\textbf{Syntax}}} & \multicolumn{4}{l}{\textit{Expression}} \\
 & \quad Comparison direction & \texttt{a < b} & \texttt{b > a} & All \\
 & \quad Operand order & \texttt{x + 1} & \texttt{1 + x} & All \\
 & \quad Redundant parentheses & \texttt{a + b * c} & \texttt{a + (b * c)} & All \\
 & \quad Length/size comparison & \texttt{len(x) > 0} & \texttt{len(x) != 0} & All \\
 & \quad Emptiness test & \texttt{not x} & \texttt{len(x) == 0} & Py/Java \\
 & \quad Default argument & \texttt{range(n)} & \texttt{range(0, n)} & Py/Java \\
 & \quad De Morgan form & \texttt{!(a \&\& b)} & \texttt{!a || !b} & Java/C\texttt{++} \\
 & \quad Membership container & \texttt{x in (a, b)} & \texttt{x in [a, b]} & Py \\
 & \quad Merged comparison & \texttt{x in (a, b)} & \texttt{x == a or x == b} & Py \\
 & \quad Power operator & \texttt{x ** y} & \texttt{pow(x, y)} & Py \\
 & \quad Slice start & \texttt{x[:n]} & \texttt{x[0:n]} & Py \\
 & \quad Reversed range & \texttt{reversed(range(n))} & \texttt{range(n-1, -1, -1)} & Py \\
 & \quad Empty list & \texttt{[]} & \texttt{list()} & Py \\
 & \quad Digit grouping & \texttt{1000000} & \texttt{1\_000\_000} & Py \\
\addlinespace[3pt]
 & \multicolumn{4}{l}{\textit{Declaration}} \\
 & \quad Augmented assignment & \texttt{x += 1} & \texttt{x = x + 1} & All \\
 & \quad Increment form & \texttt{i++} & \texttt{++i} & Java/C\texttt{++} \\
 & \quad Multiple declaration & \texttt{int a, b;} & \texttt{int a; int b;} & Java/C\texttt{++} \\
 & \quad \texttt{final}/\texttt{const} local & \texttt{int x = f();} & \texttt{final int x = f();} & Java/C\texttt{++} \\
 & \quad Tuple assignment & \texttt{a, b = x, y} & \texttt{a = x; b = y} & Py \\
 & \quad Chained assignment & \texttt{x = y = 0} & \texttt{x = 0; y = 0} & Py \\
\addlinespace[3pt]
 & \multicolumn{4}{l}{\textit{Control flow}} \\
 & \quad Branch order & \texttt{if (c) A else B} & \texttt{if (!c) B else A} & All \\
 & \quad Conditional expression & \texttt{x = c ? a : b;} & \texttt{if (c) x=a; else x=b;} & All \\
 & \quad Infinite loop & \texttt{for (;;)} & \texttt{while (true)} & All \\
 & \quad Return parentheses & \texttt{return x;} & \texttt{return (x);} & Py/Java \\
 & \quad Braces on one-statement body & \texttt{if (c) s;} & \texttt{if (c) \{ s; \}} & Java/C\texttt{++} \\
 & \quad Empty loop body & \texttt{while (c);} & \texttt{while (c) \{\}} & Java/C\texttt{++} \\
 & \quad Explicit \texttt{None} return & \texttt{return} & \texttt{return None} & Py \\
 & \quad Placeholder body & \texttt{...} & \texttt{pass} & Py \\
 & \quad List comprehension & \texttt{[f(i) for i in it]} & \textit{loop with} \texttt{append} & Py \\
 & \quad \texttt{any()} vs.\ loop & \texttt{any(c for i in it)} & \textit{loop with early return} & Py \\
\midrule
\multirow{5}{*}{\rotatebox[origin=c]{90}{\textbf{Formatting}}} & Operator spacing & \texttt{a + b} & \texttt{a+b} & All \\
 & Keyword spacing & \texttt{if\ (c)} & \texttt{if\ \ (c)} & All \\
 & Closing-bracket indent & \textit{at opening indent} & \textit{one level deeper} & All \\
 & Blank lines before a def. & \textit{as in the guide} & \textit{one fewer} & All \\
 & Final newline & \textit{present} & \textit{absent} & All \\
\bottomrule
\end{tabularx}
\end{table}

\end{document}